\documentclass[12pt]{article}

\usepackage{amssymb}
\usepackage{graphicx}
\usepackage{amsmath}
\usepackage{amsfonts}
\usepackage{multirow}
\usepackage{mathtools}
\usepackage{float}
\usepackage{a4wide}
\usepackage{yfonts}
\usepackage[justification=centering,font=small]{subcaption}
\usepackage{standalone}
\usepackage{gnuplot-lua-tikz}
\usepackage{xcolor}
\usepackage{lipsum}
\usepackage{soul}

\usepackage{tikz}
\usetikzlibrary{calc}
\usetikzlibrary{matrix}
\usetikzlibrary{arrows}
\usetikzlibrary{positioning}
\usetikzlibrary{decorations.text}

\DeclarePairedDelimiter\floor{\lfloor}{\rfloor}

\title{Cooperation and antagonism in information exchange in a growth scenario with two species}
\author{Andr\'{e}s C. Burgos$^1$ and Daniel Polani$^2$ \\
\mbox{}\\
$^{1,2}$Adaptive Systems Research Group, University of Hertfordshire, Hatfield, UK \\
$^1$Email address: a.c.burgos@herts.ac.uk, Tel: +44 1707 28 4490}

\date{}

\begin{document}
\maketitle

\begin{abstract}
We consider a simple information-theoretic model of communication, in which two species of bacteria have the option
of exchanging information about their environment, thereby improving their chances of survival. For this purpose, we model a system
consisting of two species whose dynamics in the world are modelled by a bet-hedging strategy. It is well known that such models lend
themselves to elegant information-theoretical interpretations by relating their respective long-term growth rate to the information
the individual species has about its environment. We are specifically interested in modelling how this dynamics are affected when
the species interact cooperatively or in an antagonistic way in a scenario with limited resources. For this purpose, we consider
the exchange of environmental information between the two species in the framework of a game. Our results show that a transition
from a cooperative to an antagonistic behaviour in a species results as a response to a change in the availability of resources.
Species cooperate in abundance of resources, while they behave antagonistically in scarcity.
\end{abstract}

\section{Introduction}
\label{sec:introduction}


Information is a central concept in biology. The ability of living organisms to acquire and process information about their environment
is essential for their survival and reproduction. This is particularly crucial for organisms living in fluctuating environments, where
they face the challenge of adapting to unpredictable circumstances. The failure of a species’ population to anticipate such changes
could be fatal, eventually leading the species to extinction. In environments where reliable cues that a species requires to survive are
present, sensing the environment may eliminate environmental uncertainty, allowing the species to adopt a suitable phenotype for the
current conditions. However, when uncertainty remains in the environment, a species will follow a \emph{bet-hedging} strategy
\cite{Slatkin1974, Seger1987}, where it tries to maximise its long-term growth rate by adopting different phenotypes for each of the
possible environmental conditions, in proportions based on the information about the environment they possess. The classic example of
bet-hedging in biology is Cohen's model of seed dormancy, where a seed germinates stochastically in different periods relative to the
probability of rainfalls \cite{Cohen1966}.

The relation between information and long-term growth rate was first formalized by Kelly using the example of a horse race, where a gambler
receiving side information about the race maximises its capital's long-term growth rate by betting proportionally to the updated
probabilities each horse has of winning \cite{Kelly1956}. The same principle was considered in models of evolution of biological systems
living in fluctuating environments \cite{Dempster1955, Levins1962, Cohen1966}, and the relationship between information and long-term
growth rate was analysed in information-theoretic terms in \cite{Kussell2005, Bergstrom2004, Bergstrom2005, Donaldson-Matasci2008,
MatinaC.Donaldson-Matasci2010, Rivoire2011}, where it is shown that an increase in environmental information of a species is translated
into an increase in its long-term growth rate.

Bacteria, as many other organisms living in fluctuating environments, must constantly make adaptive decisions in order to survive
\cite{Perkins2009, Balazsi2011}. For instance, bacteria have the ability to switch its phenotype to a more suited one when facing a
change in environmental conditions \cite{Elowitz2002, Balaban2004, Leisner2008, Fraser2009, Lopez2009}. The decision to adopt a particular
phenotype is based upon its information about the environment, and when the future conditions cannot be perfectly predicted, bacteria
will hedge their bets \cite{Veening2008a, Beaumont2009}. 
This stochastic decision-making process, where a cell adopts a phenotype with a certain probability, can be considered as the outcome of a
complex internal biochemical network, and therefore as an evolvable trait \cite{Tagkopoulos2008, Perkins2009, Lopez2009}. 



Besides sensing environmental factors such as temperature, oxygen, pH levels, etc., bacteria also obtain information about their
environment by detecting concentration levels of diffusable cues released by the same bacterial species or by other species of bacteria
\cite{Fuqua1994, Surette1999, Miller2001}. This process is commonly known as \emph{quorum sensing}, although the original interpretation
was more restrictive. Originally, the diffusable cues were only considered as an indicator of cell density, where a sufficiently large
concentration of these cues would indicate that a quorum of cells was achieved \cite{Fuqua1994, Surette1999}. This quorum allows bacteria
to perform diverse physiological activities such as secretion of virulence factors, formation of biofilms, conjugation, sporulation and
bioluminiscence \cite{Miller2001, Henke2004}. 

Since the introduction of quorum sensing, other uses for diffusable cues by bacteria have been found. For instance, in diffusion sensing,
bacteria employ cues to monitor diffusion in their environment \cite{Redfield2002}. Another study relates bacterial cues to pH levels
in the environment, a process called diel sensing, which, due to pH fluctuations, shows a daily cycle \cite{Decho2009}. A list of different
uses for diffusable cues by bacteria can be found in \cite{Platt2010}, where they propose to utilise the term quorum sensing to refer
to these processes, without restricting its meaning to a method of measuring cell density. Instead, the term should be considered as a
general method to indirectly obtain information about environmental factors that influence the accumulation and perception of the cues.


Considering this, we propose a theoretical model which combines the two mentioned aspects of bacteria: bet-hedging and cell-to-cell
communication, where cells exchange information about the environmental conditions on which they depend and are trying to predict.
We will neither attempt to model any particular mechanism to integrate the different sources of environmental information, nor intend
to model how a cell chooses a phenotype. Instead, we will model the dynamics of bacteria cells in a generic information-theoretic
framework, such that bacterial communication becomes an illustrative interpretation of a general model of growth with information
exchange in a scenario with limited resources. Other interpretations of the model are discussed in Sec. \ref{sec:discussion}.
Information theory \cite{Shannon1948} allows general high-level descriptions of systems, permitting to hide away irrelevant details for
the purposes of a model \cite{Polani2009, Nemenman2011}.  In particular, information theory provides a natural framework to analyse cells'
decision-making processes in uncertainty where the mechanisms need not to be modelled \cite{Mian2011, Waltermann2011, Brennan2012, Rhee2012}.

In taking this view, we focus on the emergent behaviours related to information exchange between two species of bacteria following a
bet-hedging strategy in a scenario with limited resources.
In our model, the consumption of resources as well as the amount of environmental information (from the same and from the other species)
are density-dependent. Larger populations can potentially share more environmental information than smaller ones, increasing the long-term
growth rate of recipient cells. Thus, a species can actively increase the information about the environment it could perceive in the
future, by sharing information with the other species, thereby increasing its population. On the other hand, larger populations consume
more resources, which affects the survival of a species' population, and therefore the environmental information the cells in the
population acquire. We analyse this trade-off through a game, where two species of bacterial cells competing for resources have the option 
to share all of their environmental information with the other species.

Other game-theoretical models have also considered dynamical payoffs \cite{Tomochi2002, Santos2006, Lee2011, Requejo2011, Requejo2012}.
In particular, \cite{Requejo2011, Requejo2012} considered a model based on limited resources, achieving qualitatively similar results. Both
in their work and ours, there is a transition in the dominant strategy resulting from a change in the availability of resources. This
transition is from a game equivalent to a Prisoner's Dilemma, where defection is dominant, to a Harmony Game, where cooperation dominates.
In this study, we present a model from an information-theoretic perspective.


While the majority of evolutionary game-theoretical models assume species with fixed strategies during their lifetime, and then analyse
the composition of the resulting population (cooperators vs. defectors), here we want to study which are the best \emph{communication}
strategies for a species based on the information it has about its context. Where optimal strategies for communication exist, they
would serve as an indication of which behaviours of a species evolution would favour. For other cases, we discuss possible modifications
of the model in order to study them.



\section{Model}
\label{sec:model}

\subsection{Overview}

We consider a model where two different species of bacteria can sense complementary information about their environment and have the
ability to share that information with other species. Both species follow a bet-hedging strategy, where the environmental information 
they obtain is translated into growth rate. Therefore, the more information about their environment they obtain, the higher their growth
rate will be. We want to study whether the species would communicate (cooperate) or behave antagonistically in scenarios where they
both depend on a common resource for their survival.

For this purpose, we consider a minimal model that is able to capture the communication behaviour of a species. We imagine an environment
that can be in one of four equally likely states (\emph{i.e.} its entropy is $2$ bits) and that each species can potentially sense only one
of the two bits. In this way, species depend on each other to eliminate (approximately) their environmental uncertainty, creating a mutual
interest in their survival. In addition, we assume that each individual cell can measure its corresponding bit with only 85\% accuracy.

We consider two types of communication that can help bacteria obtain more information about the environment: (a) within-species
communication, in which each member of the population can integrate completely the information from all other members of the population
(so that even though each individual can sense the species-specific environmental bit with only 85\% accuracy, several bacteria from the
same species can talk to each other to obtain close to the entire bit). Thus, the total information available to an individual increases
with population size. (b) between-species communication, in which a receiver species can incorporate all the information from the
individuals of the sender-species. As before, the amount of information that is available increases with the population size.

We assume that all the information shared, either within- or between-species, is fully interpretable by the receiver, and thus can always
be translated into growth rate (via improving their bet-hedging strategy). This is an idealisation that allow us to focus on
communication strategies rather than on the interpretation of information. Furthermore, in order to give both species the option of not
communicating with the other species, while still being able to communicate with members of its own species, we assume three different
chemical languages: each species communicate through its own particular language, which cannot be understood by the other species; and
we assume a joint language for between-species communication. All communication happens through an idealised non noisy-channel.

We want to study how scenarios with limited resources for bacteria affects their communicating behaviour. For this purpose, we consider
a game where, under the assumption that within-species communication always happen, we give the option to each species to share
their environmental information with the other species.

In the system, at each time-step, the available resources are distributed among the species such that an equal proportion of both species'
populations survive. This eliminates favouritism towards a species, assuming an unstructured distribution of resources. Then, if resources are
enough for both populations, then all of them survive. Of those that consumed resources, the ones that match the demands of the environmental
conditions further survive and reproduce. The proportion of a population that matches the environment will depend on how much information
each individual has: in the ideal case where all of them know the future environmental condition, 100\% of the population that consumed
resources survives and reproduce.

Then, if a species shares information with the other, the latter will increase its growth rate, since it will improve its prediction of
the environmental conditions. This can be beneficial for the species that shared information, since by increasing the other species' population
size, it increases the amount of environmental information that the latter can potentially share back to the former. We note here that we are
considering at least a second order process, where the consequences of a species can only be perceived at a later stage. However, increasing
the other species' growth rate have the disadvantage that resources are depleted at a faster rate. This trade-off between environmental
information and resources is what we study here through a game-theoretic scenario.

\subsection{Model introduction and outline}

\begin{figure}[ht]
\centering
  \begin{tikzpicture}
    [->,>=stealth',shorten >=2pt,auto,node distance=2cm,
    thick,main node/.style={font=\sffamily\normalsize\bfseries}]

    \node[main node] (1) [] {$E_t$};
    \node[main node] (2) [right of=1,node distance=4cm] {$E_{t + 1}$};
    \node[main node] (3) [right of=2,node distance=4cm] {$E_{t + 2}$};
    \node[main node] (4) [above right of=1] {$S_{X_t}$};
    \node[main node] (5) [right of=4,node distance=4cm] {$S_{X_{t + 1}}$};
    \node[main node] (6) [right of=5,node distance=4cm] {$S_{X_{t + 2}}$};
    \node[main node] (7) [above of=4,node distance=2cm] {$X_t$};
    \node[main node] (8) [above of=5,node distance=2cm] {$X_{t + 1}$};
    \node[main node] (9) [above of=6,node distance=2cm] {$X_{t + 2}$};
    \node[main node] (10) [below right of=1] {$S_{Y_t}$};
    \node[main node] (11) [right of=10,node distance=4cm] {$S_{Y_{t + 1}}$};
    \node[main node] (12) [right of=11,node distance=4cm] {$S_{Y_{t + 2}}$};
    \node[main node] (13) [below of=10,node distance=2cm] {$Y_t$};
    \node[main node] (14) [below of=11,node distance=2cm] {$Y_{t + 1}$};
    \node[main node] (15) [below of=12,node distance=2cm] {$Y_{t + 2}$};
    \node[main node] (17) [right of=9,node distance=3.5cm] {$\dots$};
    \node[main node] (18) [right of=15,node distance=3.5cm] {$\dots$};

    \path[every node/.style={font=\sffamily\small}]
      (1) edge node {} (4)
      (1) edge node {} (10)
      (2) edge node {} (5)
      (2) edge node {} (11)
      (3) edge node {} (6)
      (3) edge node {} (12)
      (7) edge node {$p_t$} (4)
      (8) edge node {$p_{t + 1}$} (5)
      (9) edge node {$p_{t + 2}$} (6)
      (13) edge node {$p_t$} (10)
      (14) edge node {$p_{t + 1}$} (11)
      (15) edge node {$p_{t + 2}$} (12)
      (7) edge node {$\delta_{X_t}$} (8)
      (8) edge node {$\delta_{X_{t + 1}}$} (9)
      (9) edge node {$\delta_{X_{t + 2}}$} (17)
      (13) edge node {$\delta_{Y_t}$} (14)
      (14) edge node {$\delta_{Y_{t + 1}}$} (15)
      (15) edge node {$\delta_{Y_{t + 2}}$} (18)
      ;
  \end{tikzpicture}
  \caption{\small{Bayesian network describing the relation between the main variables of the model. $E_t$ denotes the environmental conditions at time-step $t$,
				$S_{x_t}$ ($S_{y_t}$) the sensors of species $X$ ($Y$) at time-step $t$, and $X_t$ ($Y_t$) the population density of species $X$ ($Y$) at time-step $t$.}}
  \label{fig:model_dynamics}
\end{figure}
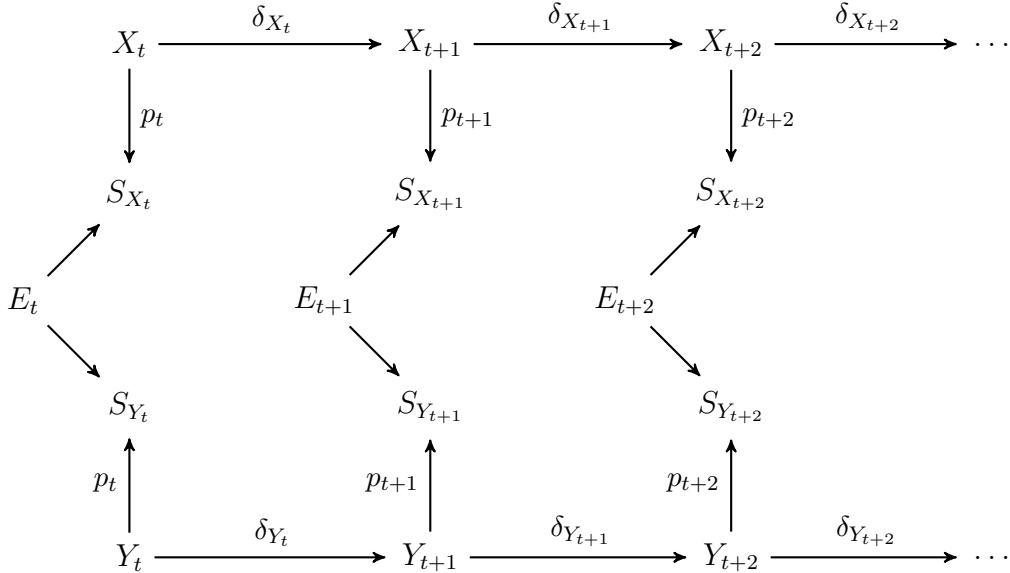

Let us give an introduction to the model, which mainly describes the relations between the variables of the system shown in
Fig. \ref{fig:model_dynamics}. Our system consists of populations of two species of bacterial cells, $X$ and $Y$,
both living in the same environment and depending on the same set of environmental conditions for survival. We assume temporarily varying
conditions, and, therefore, at each time-step, one of these environmental conditions occur. We model the environmental conditions at
time-step $t$ by a random variable $E_t$ with four states, where $p(e_t)$ is the probability of condition $e_t$ to occur. 

Each individual cell of each species acquires information about the environment through its sensors, which are denoted $S_{X_{i_t}}$ for
a cell $i$ of species $X$ at time-step $t$, and $S_{Y_{j_t}}$ for a cell $j$ of species $Y$ at time-step $t$. We denote the sensors of the
\emph{population} of species $X$ at time-step $t$ by the random variable $S_{X_t}$, and the sensors of the \emph{population} of species $Y$
at time-step $t$ by $S_{Y_t}$ (the sensor variable of a population is a function of the individual sensors, we will explain later how this
is computed).

The number of individual cells of species $X$ ($Y$) that acquire environmental information at time-step $t$ is a function of the population
density $X_t$ ($Y_t$). However, only the proportion $p_t$ of the population that consume resources (and therefore survives) is able to capture
information about the environment. When resources are not sufficient for the consumption of both populations, we assume that both population
densities are reduced by a proportionality factor $p_t$. This factor depends on the available resources (introduced later in Sec. \ref{sec:gr}),
as well as on the population densities.

Now, the rate at which species $X$ grow at time-step $t$ is given by $\delta_{X_t}$, and depends on the amount of environmental information
of the population at that time-step. This information can increase if, for instance, species $Y$ shares its environmental information with
$X$. Moreover, if species $X$ shares back information with species $Y$, then species $Y$ would have increased its growth rate by sharing
information in a previous time-step. However, larger populations consume more resources, and $Y$'s growth rate may decrease as a consequence
of sharing information. Therefore, different strategies are optimal depending on initial conditions, and this is what we will study using a
game-theoretical framework.


In the following sections we explain the model in detail. The dynamics of the population are defined in Sec. \ref{sec:population_dynamics}.
In Sec. \ref{sec:envinfo}, we explain how cells acquire information about their environment. The information is obtained from three
\emph{exclusive} sources: from environmental cues not related to bacteria (these cues infer different sensor states on bacteria, see Sec.
\ref{sec:sensoryinfo}); from communicated information from cells of the same species (within-species communication, see Sec.
\ref{sec:withincomm}); and from communicated information from cells of the other species (between-species communication, see Sec.
\ref{sec:betweencomm}). In Sec. \ref{sec:bethedging}, we show how the environmental information each cell obtains, when a bet-hedging
strategy is followed, translates into the long-term growth rate of a population. Finally, we study through a game presented in
Sec. \ref{sec:infgame} the optimal communication behaviour of the species in different scenarios.

\subsection{Environment}
\label{sec:environment}

Our system consists of populations of two species of bacterial cells, $X$ and $Y$, both living in the same environment and depending on
the same set of environmental conditions for survival. These conditions are assumed to be independent of bacterial populations, and
therefore are not affected by their consumption or production of substances in the environment. For instance, bacteria may need to adapt
its phenotype to a change in temperature, or in pH levels, sugar concentration, or any combination of environmental factors. While
the range of these variables may be in the continuum, we assume a partition of the range into relevant states for the organism survival.

We assume temporarily varying conditions, and, therefore, at each time-step, one of these environmental conditions occur. We model the
environmental conditions at time-step $t$ by a random variable $E_t$ with four states, where $p(e_t)$ is the probability of condition
$e_t$ to occur. Without loss of generality, we assume $E_t$ to be uniformly distributed. Additionally, we assume environmental conditions
to be independent and identically distributed (\emph{i.i.d.}) in time.

\subsection{Population dynamics}
\label{sec:population_dynamics}

We model the dynamics of populations of species $X$ and $Y$ by \emph{logistic maps}:

\begin{equation}
	\left\{
		\begin{array}{l}
		X_{t + 1} = \delta_{X_t}\;X_t\;(1 - X_t) \\
		Y_{t + 1} = \delta_{Y_t}\; Y_t\;(1 - Y_t) \\
		\end{array}
	\right.
	\label{eq:populations}
\end{equation}

where $X_t$ and $Y_t$ represent the population density of species $X$ and $Y$ at time $t$, respectively. The density
is the ratio of the existing population to the carrying capacity, which in our case is set to to $1$ for both populations.

The logistic map is a simple non-linear difference equation with complex behaviour, generally used in ecology and biology
to model population growth, but also used in other research areas, such as genetics, epidemiology and economics \cite{May1976}.
This equation has interesting properties that makes it attractive to use, such as proportional growth at low densities and asymptotic
growth at high densities. The value $\delta_X$ and $\delta_Y$ are the rates at which population $X$ and $Y$ grows, respectively, which
depend on the amount of environmental information each species have obtained. For non-trivial dynamical behaviour, $1 < \delta_X < 4$
and $1 < \delta_Y < 4$ is required \cite{May1976}. We will use values such that $0 < \delta_X \leq 2$ and $0 < \delta_Y \leq 2$,
assuming $2$ as a reproductive limit. To define the growth rate of a species, we first need to compute how much information about
the environment it acquires, which we do in the following section.

\subsection{Environmental information of an individual cell}
\label{sec:envinfo}

In this section we define how we compute the amount of information an individual cell obtains from each of the possible sources
we are considering: its sensors, information shared by individuals of the same species, and information shared by individuals
of the other species. We recall that an individual would obtain more environmental information when the density of the population
of the communicating species is larger. The densities of the populations are given by Eq. \ref{eq:populations}.

\subsubsection{Sensory information of individual cells}
\label{sec:sensoryinfo}

Each individual cell of both species sense cues from the environment. We represent the sensors of an individual cell $i$ of species $X$
as a random variable $S_{X_i}$, and the sensors of an individual cell $j$ of species $Y$ as a random variable $S_{Y_j}$. We define the
conditional probabilities $\Pr\left(S_{X_i}\;\middle\vert\;E\right)$ for every individual $i$ of species $X$ and
$\Pr\left(S_{Y_j}\;\middle\vert\;E\right)$ for every individual $j$ of species $Y$ in the system, and thus we can measure the amount of
information that each individual acquires from the environment by
computing the mutual information between its sensor variable and the environmental variable (see \ref{sec:inftheory} for
information-theoretic definitions). These values are bounded by the entropy of the environment, which in our case is $H(E) = 2$ bits.

In order to avoid giving an advantage to a species, we assume that all individuals of both species acquire the same amount of
environmental information (this becomes important later when this amount is translated into long-term growth rate). Nevertheless,
the difference between the two species is in the aspects of environmental information that they capture, as shown in the conditional
probabilities Eq. \ref{eq:condprobx} and Eq. \ref{eq:condproby}. Individuals of species $X$ capture information only about two states
of the environment $E$, being unable to sense the other two states. On the contrary, individuals of species $Y$ capture information
only about the two states species $X$ cannot sense, while being unable to sense the other two states. The amount of information about
the environment that an individual $i$ of species $X$ captures is $I(E\;;\;S_{X_i}) = 0.39016$ bits, the same amount as an individual $j$
of species $Y$, $I(E\;;\;S_{Y_j}) = 0.39016$ bits, although the intersection of the information each of them capture is
$I(S_{X_i}\;;\;S_{Y_j}) = 0$ bits. We explain in the next sections how this assumption influences the total environmental information an
individual cell can acquire, while we analyse how it affects the results obtained from our model in \ref{sec:params}.

\begin{figure*}[ht]
	\centering
	\begin{minipage}[b]{.49\textwidth}
		\begin{equation}
		\Pr\left(S_{X_i} \;\middle\vert\; E\right) \coloneqq 
		\bordermatrix{~ & s_1 & s_2 \cr
		              e_1 & 0.85 & 0.15 \cr
		              e_2 & 0.85 & 0.15 \cr
		              e_3 & 0.15 & 0.85 \cr
		              e_4 & 0.15 & 0.85 \cr}
		\label{eq:condprobx}
		\end{equation}
	\end{minipage}\hfill
	\begin{minipage}[b]{.49\textwidth}
		\begin{equation}
		\Pr\left(S_{Y_j} \;\middle\vert\; E\right) \coloneqq 
		\bordermatrix{~ & s_1 & s_2 \cr
		              e_1 & 0.85 & 0.15 \cr
		              e_2 & 0.15 & 0.85 \cr
		              e_3 & 0.85 & 0.15 \cr
		              e_4 & 0.15 & 0.85 \cr}
		\label{eq:condproby}
		\end{equation}
	\end{minipage}
\end{figure*}

\subsubsection{Environmental information of an individual cell obtained from within-species communication}
\label{sec:withincomm}

Let us consider an individual cell $i$ of bacterial species $X$. The amount of information, on average, this cell obtains from its sensors
is $I(E\;;\;S_{X_i})$. Now, if another cell $j \not= i$ of species $X$ communicates with its own species' population (as, for instance,
by releasing a molecule into the extracellular environment), then, assuming cells of the same species share the same language, the
information about the environment of species $i$ increases (as well as that of the rest of the population). The total amount of information
about the environment for an individual cell $i$ of species $X$ when another cell $j \not= i$ communicates information is, on average,
$I(E\;;\;S_{X_i},S_{X_j})$, and the increase in environmental information for cell $i$ is $I\left(E\;;\;S_{X_j}\;\middle\vert\;S_{X_i}\right)$
(see Fig. \ref{fig:envinfodiag}).

The assumption about a common language that perfectly conveys the sensory state of a cell is an important one, and it allow us to simplify
the model by ignoring the relationship between sensor states and output signals (which are implicitly assumed to be one-to-one in this model).
A further important assumption we make regarding the population structure is that all cells perceive what other cells communicate.

As more cells communicate, the environmental information of all cells increases as shown in Fig. \ref{fig:envinfo} (see label
$I(E\;;\;S_{X_1},\dots,S_{X_n})$), considering a carrying capacity of $N = 15$ cells. In the same way, the amount of environmental information
of cells of species $Y$ increases with each individual exactly as it does in species $X$ (see label $I(E\;;\;S_{Y_1},\dots,S_{Y_m})$), with the
same carrying capacity $M$ as species $X$, $M = 15$ (this choice of value is discussed in \ref{sec:params}). In Fig. \ref{fig:envinfodiag},
we show in a different way how the environmental information of individuals increases when there is within-species communication. Each
species captures exclusive bits of environmental information, and thus when individuals of the same species communicate, they can only
reduce the uncertainty of one bit of environmental information.

Note that the computational complexity of the mutual information grows exponentially with the number of individual cells communicating
information. Since each individual cell can be in two states, the total number of states of the whole population is $2^{15}$ states.
However, we can take advantage of the fact that the probability distribution $\Pr\left(S_{X_i}\;\middle\vert\;E\right)$ is the same for
any $X_i \in [1,15]$, and thus the probability of a particular state of the population depend only on the frequencies of the states of the
individuals. In this way, the total number of possible states of a population grows linearly with population size, and we can represent the
states of a population conditioned on environmental conditions more efficiently, as we explain in \ref{sec:mettyp}. The same reasoning is
also valid for representing the population of species $Y$. The choice of $15$ as the carrying capacity of both populations was made in order
to reduce computational costs. 

Finally, since both species live in the same niche, we will assume exclusive means of communication for the species, \emph{i.e.} the
chemical language used by species $X$ and $Y$ has no overlap. Nevertheless, we consider a common language for between-species
communication. Scenarios where bacteria use different chemical languages for within- or between-species communication are common. For
instance, the bacterium \emph{Vibrio harveyi} uses two different autoinducer signals to regulate light production and other target outputs;
one mediating within-species communication, and the other between-species communication \cite{Federle2003}.

\colorlet{circle edge}{blue!50}
\colorlet{circle area}{black}

\tikzset{filled/.style={fill=circle area, draw=circle edge, thick},
    outline/.style={draw=circle edge, thick}}

\setlength{\parskip}{5mm}

\begin{figure}[ht]
	\centering
	\begin{tikzpicture}
		\filldraw[fill=gray!20] (-2,-2) rectangle (8,3.5);
	    \node[anchor=south] at (0.5,2.8) {\emph{first bit of $E$}};
	    \node[anchor=south] at (5.5,2.8) {\emph{second bit of $E$}};
		\draw [dashed] (3,3.5) -- (3,-2);
	    \draw[outline] {(0.5,1.3) circle (1.5cm)} [color=red!50, text=black, fill=red!50,font=\tiny] node at (0.5,2.35) {$I(E;S_{X_3}|S_{X_1},S_{X_2})$};
		\draw[outline] {(0.5,0.6) circle (1.6cm)} [color=red!40, fill=red!40, text=black,font=\tiny] node at (0.5,1.7) {$I(E;S_{X_2}|S_{X_1})$};
	    \draw[outline] {(0.5,-0.1) circle (1.6cm)} [color=red!30, fill=red!30, text=black,font=\footnotesize] node {$I(E;S_{X_1})$};
	    \draw[outline] {(5.5,1.3) circle (1.5cm)} [color=blue!50, text=black, fill=blue!50,font=\tiny] node at (5.5,2.35) {$I(E;S_{Y_3}|S_{Y_1},S_{Y_2})$};
	    \draw[outline] {(5.5,0.6) circle (1.6cm)} [color=blue!40, fill=blue!40, text=black,font=\tiny] node at (5.5,1.7) {$I(E;S_{Y_2}|S_{Y_1})$};
	    \draw[outline] {(5.5,-0.1) circle (1.6cm)} [color=blue!30, text=black, fill=blue!30,font=\footnotesize] node {$I(E;S_{Y_1})$};
		\node at (3, 3.9) {\emph{environmental information $E$}};
	\end{tikzpicture}
	\caption{Diagram sketching the environmental information each species captures, and how this varies when there is within-species communication. Individuals of species $X$
			capture information only about the first bit of $E$, while individuals of species $Y$ capture information only about the second bit of $E$. When, for instance,
			individuals of species $X$ communicate with each other, their environmental information increases, but it only eliminates uncertainty about the first bit. In the
			same way, individuals of species $Y$ communicating with each other can only increase their environmental information about the second bit of $E$.}
	\label{fig:envinfodiag}
\end{figure}
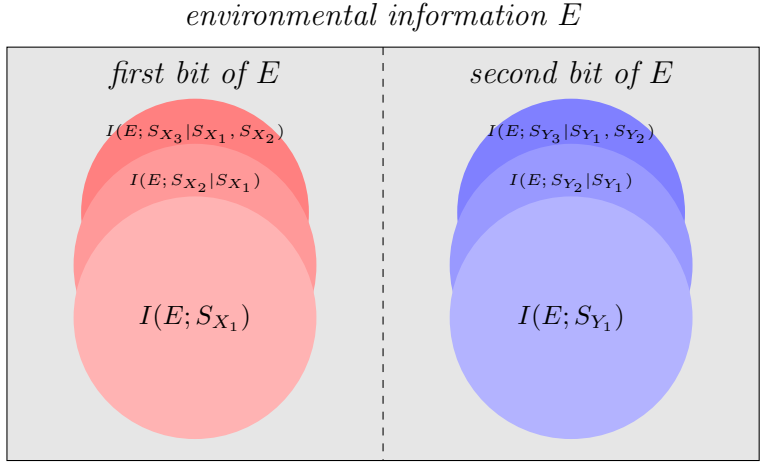


\subsubsection{Environmental information of an individual cell obtained from between-species communication}
\label{sec:betweencomm}

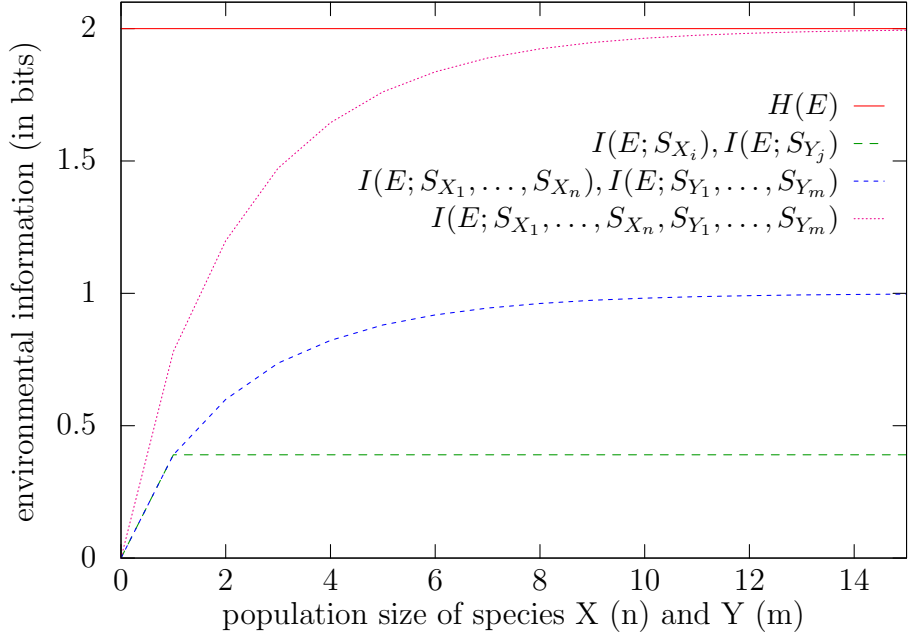
\begin{figure}[ht]
	\begin{center}
	\begin{tikzpicture}[gnuplot]
	\gpcolor{gp lt color border}
	\gpsetlinetype{gp lt border}
	\gpsetlinewidth{1.00}
	\draw[gp path] (1.504,0.985)--(1.684,0.985);
	\draw[gp path] (11.947,0.985)--(11.767,0.985);
	\node[gp node right] at (1.320,0.985) { 0};
	\draw[gp path] (1.504,2.746)--(1.684,2.746);
	\draw[gp path] (11.947,2.746)--(11.767,2.746);
	\node[gp node right] at (1.320,2.746) { 0.5};
	\draw[gp path] (1.504,4.507)--(1.684,4.507);
	\draw[gp path] (11.947,4.507)--(11.767,4.507);
	\node[gp node right] at (1.320,4.507) { 1};
	\draw[gp path] (1.504,6.268)--(1.684,6.268);
	\draw[gp path] (11.947,6.268)--(11.767,6.268);
	\node[gp node right] at (1.320,6.268) { 1.5};
	\draw[gp path] (1.504,8.029)--(1.684,8.029);
	\draw[gp path] (11.947,8.029)--(11.767,8.029);
	\node[gp node right] at (1.320,8.029) { 2};
	\draw[gp path] (1.504,0.985)--(1.504,1.165);
	\draw[gp path] (1.504,8.381)--(1.504,8.201);
	\node[gp node center] at (1.504,0.677) { 0};
	\draw[gp path] (2.896,0.985)--(2.896,1.165);
	\draw[gp path] (2.896,8.381)--(2.896,8.201);
	\node[gp node center] at (2.896,0.677) { 2};
	\draw[gp path] (4.289,0.985)--(4.289,1.165);
	\draw[gp path] (4.289,8.381)--(4.289,8.201);
	\node[gp node center] at (4.289,0.677) { 4};
	\draw[gp path] (5.681,0.985)--(5.681,1.165);
	\draw[gp path] (5.681,8.381)--(5.681,8.201);
	\node[gp node center] at (5.681,0.677) { 6};
	\draw[gp path] (7.074,0.985)--(7.074,1.165);
	\draw[gp path] (7.074,8.381)--(7.074,8.201);
	\node[gp node center] at (7.074,0.677) { 8};
	\draw[gp path] (8.466,0.985)--(8.466,1.165);
	\draw[gp path] (8.466,8.381)--(8.466,8.201);
	\node[gp node center] at (8.466,0.677) { 10};
	\draw[gp path] (9.858,0.985)--(9.858,1.165);
	\draw[gp path] (9.858,8.381)--(9.858,8.201);
	\node[gp node center] at (9.858,0.677) { 12};
	\draw[gp path] (11.251,0.985)--(11.251,1.165);
	\draw[gp path] (11.251,8.381)--(11.251,8.201);
	\node[gp node center] at (11.251,0.677) { 14};
	\draw[gp path] (1.504,8.381)--(1.504,0.985)--(11.947,0.985)--(11.947,8.381)--cycle;
	\node[gp node center,rotate=-270] at (0.246,4.683) {environmental information (in bits)};
	\node[gp node center] at (6.725,0.215) {population size of species X (n) and Y (m)};
	\node[gp node right] at (11.2,7) {\small{$H(E)$}};
	\gpcolor{gp lt color 0}
	\gpsetlinetype{gp lt plot 0}
	\draw[gp path] (11.25,7)--(11.65,7);
	\draw[gp path] (1.504,8.029)--(2.200,8.029)--(2.896,8.029)--(3.593,8.029)--(4.289,8.029)%
	  --(4.985,8.029)--(5.681,8.029)--(6.377,8.029)--(7.074,8.029)--(7.770,8.029)--(8.466,8.029)%
	  --(9.162,8.029)--(9.858,8.029)--(10.555,8.029)--(11.251,8.029)--(11.947,8.029);
	\gpcolor{gp lt color border}
	\node[gp node right] at (11.2,6.5) {\small{$I(E;S_{X_i}), I(E;S_{Y_j})$}};
	\gpcolor{gp lt color 1}
	\gpsetlinetype{gp lt plot 1}
	\draw[gp path] (11.25,6.5)--(11.65,6.5);
	\draw[gp path] (1.504,0.985)--(2.200,2.359)--(2.896,2.359)--(3.593,2.359)--(4.289,2.359)%
	  --(4.985,2.359)--(5.681,2.359)--(6.377,2.359)--(7.074,2.359)--(7.770,2.359)--(8.466,2.359)%
	  --(9.162,2.359)--(9.858,2.359)--(10.555,2.359)--(11.251,2.359)--(11.947,2.359);
	\gpcolor{gp lt color border}
	\node[gp node right] at (11.2,6) {\small{$I(E;S_{X_1},\dots,S_{X_n}),I(E;S_{Y_1},\dots,S_{Y_m})$}};
	\gpcolor{gp lt color 2}
	\gpsetlinetype{gp lt plot 2}
	\draw[gp path] (11.25,6)--(11.65,6);
	\draw[gp path] (1.504,0.985)--(2.200,2.359)--(2.896,3.096)--(3.593,3.579)--(4.289,3.881)%
	  --(4.985,4.086)--(5.681,4.220)--(6.377,4.311)--(7.074,4.372)--(7.770,4.415)--(8.466,4.443)%
	  --(9.162,4.463)--(9.858,4.476)--(10.555,4.486)--(11.251,4.492)--(11.947,4.497);
	\gpcolor{gp lt color border}
	\node[gp node right] at (11.2,5.5) {\small{$I(E;S_{X_1},\dots,S_{X_n},S_{Y_1},\dots,S_{Y_m})$}};
	\gpcolor{gp lt color 3}
	\gpsetlinetype{gp lt plot 3}
	\draw[gp path] (11.25,5.5)--(11.65,5.5);
	\draw[gp path] (1.504,0.985)--(2.200,3.733)--(2.896,5.207)--(3.593,6.173)--(4.289,6.778)%
	  --(4.985,7.187)--(5.681,7.454)--(6.377,7.638)--(7.074,7.760)--(7.770,7.844)--(8.466,7.901)%
	  --(9.162,7.941)--(9.858,7.967)--(10.555,7.986)--(11.251,7.999)--(11.947,8.008);
	\gpcolor{gp lt color border}
	\gpsetlinetype{gp lt border}
	\draw[gp path] (1.504,8.381)--(1.504,0.985)--(11.947,0.985)--(11.947,8.381)--cycle;
	\gpdefrectangularnode{gp plot 1}{\pgfpoint{1.504cm}{0.985cm}}{\pgfpoint{11.947cm}{8.381cm}}
	\end{tikzpicture}
	\caption{Total amount of environmental information for different scenarios: $H(E)$ is the uncertainty of the environment,
			$I(E;S_{X_i})$ and $I(E;S_{Y_j})$ correspond to the case where an individual cell $i$ of species $X$ and an individual cell
			$j$ of species $Y$ acquire information from their sensors only, respectively. $I(E;S_{X_1},\dots,S_{X_n})$ is the total amount
			of information of each cell of species $X$ when $n$ cells communicate; in the same way $I(E;S_{Y_1},\dots,S_{Y_m})$ is the total
			amount of information of each cell of species $Y$ when $m$ cells communicate. $I(E;S_{X_1},\dots,S_{X_n},S_{Y_1},\dots,S_{Y_m})$
			is the total amount of environmental information each cell of both population have when $n$ cells of species $X$ and $m$ cells of
			species $Y$ communicate.
			}
	\label{fig:envinfo}
	\end{center}
\end{figure}

In the same way as last section, individual cells also obtain information from cells of the other species. Again, we need to assume
a common code between the species. However, as stated before, the chemical language used for between-species communication need to
be different from both of the within-species communication languages. Then, an individual cell $i$ of species $X$ acquiring communicated
environmental information from cell $j$ of species $X$ and cell $k$ from species $Y$ will have an amount equal to
$I(E;S_{X_i},S_{X_j},S_{Y_k})$. In Fig. \ref{fig:envinfo} we show the amounts of environmental information an individual cell acquires
in different scenarios: with only sensory information, with within-species communication and with between-species communication.

Let us note that, since each species is specialised to capture different aspects of the environment, the contribution (from a cell's
perspective) of a first cell of the other species is significantly higher (in terms of environmental information) than that of a cell of
the same species. This can be appreciated in Fig. \ref{fig:envinfodiag}: if we consider an individual cell $1$ of species $X$, its
environmental information is $I(E;S_{X_1}) = 0.39016$ bits. If cell $1$ of species $Y$ communicates information, then the total amount of
information for cell $1$ of species $X$ is $I(E;S_{X_1},S_{Y_1}) = I(E;S_{X_1}) + I(E;S_{Y_1}) = 0.78032$ bits (since $I(S_{X_1};S_{Y_1}) = 0$
bits); while if a cell of the same species shares information, the increase in environmental information is $I(E;S_{X_2}|S_{X_1}) = 0.209267$
bits, in the same way that if another cell $2$ of the other species share information, the increase is $I(E;S_{Y_2}|S_{Y_1}) = 0.209267$ bits.

\subsection{Bet-hedging on environmental conditions}
\label{sec:bethedging}

\subsubsection{Long-term growth rate of a bacterial population}
\label{sec:ltgr}

Bacterial cells in our system adopt one of a set of possible phenotypes at each time-step. For each possible environmental condition, we
assume there is only one phenotype that meets its demands and thus allows the cell to survive. Then, cells adopting a phenotype other
than the one that meets the demands of the current environmental conditions die out. This simplifying assumption will allow us to
express the relationship between environmental information and long-term growth rate in a more elegant way. We explain in Sec.
\ref{sec:ltgr} the consequences to our model of removing this assumption. Without loss of generality, we assume that the environmental
conditions and the phenotypes are labelled from the set $\{ 1, 2, 3, 4 \}$. We define the \emph{reproduction rate} $f$ of a bacterial cell
adopting phenotype $\varphi$ when the environmental condition $e_t$ occur as the following function:

\begin{equation}
f(\varphi, e_t) = \left\{
	\begin{array}{rl}
		2 &\mbox{ if $\varphi = e_t$} \\
		0 &\mbox{ otherwise}
	\end{array}
	\right.
	\label{eq:ures}
\end{equation}

Bacterial cells are complex organisms, with intricate biochemical networks. As it is recognised in several studies, these internal
networks in bacteria enable predictive behaviour in a probabilistic fashion \cite{Libby2007, Tagkopoulos2008, Perkins2009}. Then,
individual cells will develop one of its possible phenotypes with some probability, and we regard the probability distribution over the
phenotypes as the \emph{betting strategy} $\pi$ of an individual cell. Which strategy do cells follow?

As it is shown in theoretical \cite{MatinaC.Donaldson-Matasci2010} as well as empirical \cite{Beaumont2009} studies, the optimal strategy
is achieved by the proportional betting scheme $\pi = p$ (see \cite{MatinaC.Donaldson-Matasci2010, Cover2002} for a full treatment). Using a
notation similar to that used in \cite{Cover2002}, the optimal long-term growth rate of a species depending on conditions $E$ with
reproduction rate $f$ is given by:

\begin{equation}
W_f^\ast(E) = F - H(E) \label{eq:optgr} \\
\end{equation}



Here, $F = \sum\limits_{e} p(e) \log f(e, e)$. When a species acquires extra information $C$ about the environment, then the
long-term growth rate is given by:

\begin{align}
W_f^\ast\left(E\;\middle\vert\;C\right) &= F - H\left(E\;\middle\vert\;C\right) \label{eq:we} \\
	&= F - H(E) + I(E\;;\;C) \label{eq:drmi}
\end{align}

%

Equation \ref{eq:we} shows the optimal long-term growth rate for populations living in environmental conditions $E$ perceiving environmental
cues $C$. $F$ is an upper bound given by the expected reproduction rate. $H\left(E\;\middle\vert\;C\right)$ is the remaining environmental uncertainty
of each cell of the population given environmental cues $C$. Equation \ref{eq:drmi} shows the value of environmental cues $C$ in the
long-term growth rate, namely $I(E\;;\;C)$. However, if there is at least one phenotype meeting the demands of more than one environmental
conditions, then the value of $C$ in the long-term growth rate when species follow a bet-hedging strategy is not exactly $I(E\;;\;C)$, but
it is bounded by this value \cite{MatinaC.Donaldson-Matasci2010}. Moreover, for some non-diagonal functions of the reproduction rate
$f$, bet-hedging is not the optimal strategy that maximises the long-term growth rate \cite{MatinaC.Donaldson-Matasci2010}.
Therefore, our assumption that there is only one phenotype in each species that meet the demands of each of the environmental conditions
and thus survives to reproduce allows us to provide the optimal betting strategy for any case, also allowing a clear expression of
the increase in long-term growth rate. 

As a final remark, the value $F$ always equals $1$, and the growth rate of a population when it has no environmental information is
$2^{F - H(E)} = 2^{1 - 2} = 1/2$, which means that $1/4$ of the population survives and reproduces, which is the probability for an individual
cell to develop a suitable phenotype by choosing one randomly. On the other hand, when the uncertainty of the environment is eliminated, let us
assume by the perception of environmental cues $C$, then the growth rate of a population is $2^{F - H(E|C)} = 2^{1 - 0} = 2$, \emph{i.e.} the
whole population survives and reproduces, since they are all able to perfectly predict future environmental conditions.

\subsubsection{Growth rate of a bacterial population per time-step}
\label{sec:gr}

We now define the growth rate $\delta_{X_t}$ of a species $X$ at time-step $t$, and the growth rate $\delta_{Y_t}$ of species $Y$ at
time-step $t$. Instead of computing the growth rate of a species in one particular environment, we consider the average growth over all
possible environments. The growth rate at time-step $t$ when a species considers the information $C$ is given by:

\begin{equation}
	\delta_t \coloneqq 2^{W_f^\ast\left(E \;\middle\vert\; C\right)}
\end{equation}

For instance, the growth rate of species $X$ when $n$ individuals share information only within their species is
$\delta_{X_t} = 2^{W_f^\ast(E|S_{X_{1_t}},\dots,S_{X_{n_t}})}$. In a similar way, the long-term growth in one time-step of species
$Y$ when $m$ individuals share information only within their species is $\delta_{Y_t} = 2^{W_f^\ast(E|S_{Y_{1_t}},\dots,S_{Y_{m_t}})}$.
Here, $n$ needs to be related to $X_t$ (the current population density of species $X$) and $m$ needs to be related to $Y_t$
(the current population density of species $Y$).

In our model, population densities are represented by real values in the range $[0,1]$, and we need to map this range to a number
of individual cells to able to compute the long-term growth rate, which requires computing values $I(E;S_{X_{1_t}},\dots,S_{X_{n_t}})$
for $n_t$ individuals of species $X$ and $I(E;S_{Y_{1_t}},\dots,S_{Y_{m_t}})$ for $m_t$ individuals of species $Y$. However, only the
individuals that are able to consume resources (and therefore survive) sense the environment. The proportion of individuals that consume
resources at time-step $t$ is given by

\begin{equation}
p_t \coloneqq \left\{
	\begin{array}{rl}
		1 &\mbox{ if $R_t > X_t + Y_t$} \\
		\frac{R_t}{X_t + Y_t} &\mbox{ otherwise}
	\end{array}
	\right.
	\label{eq:proportion}
\end{equation}

In Eq. \ref{eq:proportion}, both populations survive if resources are sufficient for their consumption. However, when they are not
sufficient, the proportion of each population that survives is proportional to the ratio of resources to the sum of the population densities.
The dynamics of the resources is defined as follows:

\begin{equation}
	R_{t + 1} \coloneqq
	\left\{
		\begin{array}{rl}
		\alpha \big(R_t - (X_t + Y_t)\big) &\mbox{ if $R_t - (X_t + Y_t) > 0$} \\
		0 &\mbox{ otherwise}
		\end{array}
	\right.
	\label{eq:resources}
\end{equation}

In Eq. \ref{eq:resources}, the resources are depleted relative to the population densities, and, if there are any left, they grow by a
factor $\alpha$. Once resources are depleted, they remain in that state. In appendix \ref{eq:resources_replenishment}, we consider
resources that are periodically replenished instead of the dynamics described above.

Now we can compute the number of individuals that sense the environment at time-step $t$, which is given by $n_t = p_t \times X_t \times N$
for species $X$ and $m_t = p_t \times Y_t \times M$ for species $Y$ ($N$ and $M$ are the assumed carrying capacity for the population of
species $X$ and $Y$, respectively). As stated earlier, when $n_t$ and $m_t$ are integers, we can represent the conditional probabilities for
the populations, $\Pr(S_{X_t}|E)$ and $\Pr(S_{Y_t}|E)$, as explained in \ref{sec:mettyp}. However, when one or both of these values are
not integers, we represent the conditional probabilities $\Pr(S_{X_t}|E)$ and $\Pr(S_{Y_t}|E)$ by interpolating between $\floor{n_t}$ and
$\floor{n_t + 1}$ individuals for species $X$ and between $\floor{m_t}$ and $\floor{m_t + 1}$ individuals for species $Y$. How we
interpolate is explained in detail in \ref{sec:interpolation}. Ideally, we would define a higher carrying capacity for both
populations (instead of $N = M = 15$, the value we use in our simulations), and then we would not need to interpolate values. However, since
computation costs grow exponentially, we overcome this difficulty by defining a small carrying capacity and simulating ``intermediate''
sizes of the population. In any case, what matters in our model is the amount of environmental information of each species, rather than
the actual amount of individuals composing the population.

\subsection{Game between the species}
\label{sec:infgame}

\subsubsection{Introduction}

In order to study the communication behaviour of the species, we set up a game where they can either share information (cooperate),
or behave antagonistically. The goal of each species is to \textbf{maximise their growth rate for a local look-ahead}. We explain
below why species need a look-ahead for making decisions related to communication, and how this relates to species that bet to
maximise their long-term growth rate.

The game we propose here differs from traditional evolutionary game theory in that, in our model, an organism does not have an
inherited (and fixed during its lifetime) strategy. Instead, we consider species whose communication behaviour depends and changes on
the context.

The context of the species is composed of several variables: population densities, resource concentration, and other environmental
conditions. How these variables change in time is determined by the system's dynamics. Our aim is to find optimal strategies for
\emph{communication} in different contexts, and these would serve as an indication of which communication behaviours of a species
evolution would favour.

There are two important assumptions in this approach that we take: first, the species \emph{have complete knowledge} of the current
population densities and resource concentration, but their knowledge about other environmental conditions (which are relevant for their
survival) depends on information that is communicated by the same/other species. Ideally, we would consider all of these factors as
environmental information that a species needs to obtain by communicating (for instance, quorum sensing obtains densities estimates),
but here we do not -- only the survival-relevant environment state is assumed unknown to permit application of the Kelly-gambling model.

Second, the species is capable of processing the contextual information. Having complete knowledge of the former mentioned variables is
not sufficient for an organism to make a decision regarding its communication behaviour. A species would need some mechanism
(\emph{e.g.} epigenetic mechanisms) that functions as a model for the dynamics of the system the species inhabits. We argue below that
in order for bacteria to perceive the effects of their actions (sharing or not sharing information) in the rest of the system, at least
a two-step look-ahead is necessary. In other words, the mechanism needs to be a second-order process.

We are not modelling this mechanism in this paper. Instead, we are using the system's dynamics as a best case scenario for the species
to make decisions. Of course, bacteria would \emph{not} have such a detailed internal mechanism, it would be a simplified model of the
dynamics that is sufficient for predicting variables of interest. However, the system's dynamics sets the limit of what is achievable in
terms of optimal decisions, and when a species have less information about the relevant variables, their decision-making will necessarily
be worse.

Finally, in the system's dynamics, species follow a bet-hedging strategy that maximises the long-term growth rate. In the game, however,
species maximise their growth rate for a defined look-ahead. There is no conflict with these assumptions: the former is a strategy
related to how species bet on environmental conditions; the latter is about whether a species should share information or not.
Independently of the look-ahead we are using, and under the current assumptions, species always do proportional betting.

\subsubsection{Species' look-ahead}
\label{sec:look_ahead}

For this study, we assume both species consider only a local look-ahead, and thus the horizon we will be considering in the computations
is intentionally finite. For any given initial conditions $R_t$, $X_t$ and $Y_t$, the growth rate of a species' population
at that time-step depends on the population's sensory information together with the shared information from the other species (which depends
on the other species' communication strategy). Therefore, any strategy a species may take (whether it shares information with the other species
or not) would not influence its immediate payoff (\emph{i.e.} its growth rate), and hence the model does not provide an insight into
how communication strategies interact with evolution. On the other hand, if we consider species with foresight, then their strategies would
indirectly affect their payoffs, and here it then makes sense to analyse whether a species would share information or not.

A species that shares all of their environmental information would increase the growth rate of the other species. For instance,
if species $Y$ shares all the information it has available with species $X$ at time-step $t$, then $W_f^\ast\left(E_t\;\middle\vert\;S_{X_t},S_{Y_t}\right)
= F_t - H(E_t) + I(E_t\;;\;S_{X_t}) + I\left(E_t\;;\;S_{Y_t}\;\middle\vert\;S_{X_t}\right)$; if it does not share any information, then
$W_f^\ast\left(E_t\;\middle\vert\;S_{X_t}\right) = F_t - H(E_t) + I(E_t\;;\;S_{X_t})$, which is clearly less or equal than the former value.
Now, if we consider a species that maximises their growth rate at the next time-step, $t + 1$, then a species' payoff is, let us say for
species $X$ when species $Y$ shares information at time-step $t + 1$:

\begin{equation}
W_f^\ast\left(E_{t + 1}\;\middle\vert\;S_{X_{t + 1}},S_{Y_{t + 1}}\right) =
	F_{t + 1} - H(E_{t + 1}) + I(E_{t + 1}\;;\;S_{X_{t + 1}}) + I\left(E_{t + 1}\;;\;S_{Y_{t + 1}}\;\middle\vert\;S_{X_{t + 1}}\right)
\label{eq:drh1}
\end{equation}

In this equation, the terms $S_{X_{t + 1}}$ and $S_{Y_{t + 1}}$ both depend on species $X$ and $Y$'s decisions at time-step $t$, and on species
$Y$'s decision at time-step $t + 1$, but \emph{not} on species $X$'s decision at time-step $t + 1$. Let us note $W_{X_t}^\ast$ as the 
growth rate of species $X$ at time-step $t$. While we could consider $W_{X_{t + 1}}^\ast$ as the value to maximise by species $X$, it will not
reflect the consequences of the decision taken by species $X$ at time-step $t + 1$. In other words, this value will be always equal for different
sequences of actions (those where $X$ shares in the last time-step, and those where $X$ do not share in the last time-step).

On the other hand, if we consider longer (finite) decision horizons, we will incur into the same problem: the last action does not affect one's
payoff. For this reason, at time-step $t + horizon$, we consider the payoff to be the growth rate when the other species do not
share information, in a worst-case scenario for the species. In this way, all of the actions of both species influence the payoffs. Since in this
study we are considering a second-order process, the minimal horizon that would show any interesting behaviour in the communication strategies is
$2$. For economy in the computations and simplicity, we will use this value for the horizon.

\subsubsection{Payoff matrix}
\label{sec:payoff_matrix}

The payoff for species $X$ is given by

\begin{equation}
W_X^\ast(E_{t + 2}|S_{X_{t + 2}}) = F_{t + 2} - H(E_{t + 2}) + I(E_{t + 2};S_{X_{t + 2}}) 
\label{eq:drh2}
\end{equation}

In Eq. \ref{eq:drh2}, the values of $F_{t + 2}$ and $H(E_{t + 2})$ are fixed. The value of $S_{X_{t + 2}}$, however, depends on all the previous
decisions taken by both species. Therefore, the payoff matrix will be composed of $16$ values, since we are using a look-ahead equal to
$2$.  This payoff matrix corresponds to the
most accurate information a species could have to make a decision regarding whether it should share information or not. It is the
most accurate because it is obtained from the model itself, instead of from an organism's internal approximation. An example of a
payoff matrix is shown in Box 1.

To get an intuition on how the game will be played, we can imagine two opposite situations: first, with abundant resources, if a species shares
information in the first time-step, it will help the other species to improve their predictions, and then the collective information of a larger
population of the latter species may be ``fed back'' into the former species. We should note that, since resources are abundant, there is no damage
for a species to share information, even if the other species do not share back.
In the other case, we consider scarce resources, and then sharing information has two opposite effects: first, it increases the potential
information that can be shared back, as we improve the other species' predictions on the environment, but it also decreases the total amount of
available resources, which affects the information that both species capture. This trade-off between resources and environmental information is
what we analyse in the following section. 

\section{Results}
\label{sec:results}

We analysed the resulting payoff matrix described above for $250\times250\times300$ initial values (contexts) uniformly distributed in $[0,1]\times[0,1]
\times[0,3]$ (the range of population density of species $X$ times the range of population density of species $Y$
times the range of resources' values). We look in these matrices for dominant strategies for species $X$ (see Box 1 for strategic dominance
definitions). The parameters used (those which were not yet defined) are $\alpha = 1.05, N = M = 15$. In appendix \ref{sec:params} we discuss the
sensitivity of the parameters and the generality of the results obtained. In appendix \ref{sec:other_resource_dynamics} we consider different
dynamics for the resources in our simulations, where resources are replenished periodically instead of growing by a factor $\alpha$.

\vspace{0.5cm}
\noindent\colorbox{blue!20}{\parbox{\textwidth}{%
  \vskip10pt
  \leftskip10pt\rightskip10pt
	\footnotesize{
	\textbf{Box 1. Strategic dominance definitions.} \\
	\noindent\rule[0.5ex]{\linewidth-0.7cm}{1pt}
	The payoff matrix consists of $16$ values, where each value corresponds to the growth rate of species $X$. Each value
	results from the decisions of each species of sharing or not their environmental information in two time-steps. Below is an example of
	a payoff matrix resulting in not sharing information being strictly dominant (see definition below) for species $X$, with initial values
	$R = 1.0$, $X = 0.304$ and $Y = 0.392$.

	\vspace{0.3cm}

	\begin{tikzpicture}
	
	\matrix[matrix of math nodes,every node/.style={align=center,text width=2.5cm},row sep=0.08cm,column sep=0.50cm, nodes in empty cells,ampersand replacement=\&] (m)
	{
	-0.99890773\&-0.99907912\&-0.99889800\&-0.99907118\\
	-0.99911144\&-0.99926619\&-0.99910257\&-0.99925910\\
	-0.99891489\&-0.99908596\&-0.99890519\&-0.99907805\\
	-0.99911738\&-0.99927174\&-0.99910854\&-0.99926468\\
	};

	\draw (m.north east) rectangle (m.south west);
	\draw (m.north) -- (m.south);
	\draw (m.east) -- (m.west);

	\draw ($(m.north west)!0.25!(m.north east)$) -- ($(m.south west)!0.25!(m.south east)$);
	\draw ($(m.north west)!0.75!(m.north east)$) -- ($(m.south west)!0.75!(m.south east)$);

	\draw ($(m.north west)!0.25!(m.south west)$) -- ($(m.north east)!0.25!(m.south east)$);
	\draw ($(m.north west)!0.75!(m.south west)$) -- ($(m.north east)!0.75!(m.south east)$);

	\coordinate (a) at ($(m.north west)!0.125!(m.north east)$);
	\coordinate (b) at ($(m.north west)!0.375!(m.north east)$);
	\coordinate (c) at ($(m.north west)!0.625!(m.north east)$);
	\coordinate (d) at ($(m.north west)!0.875!(m.north east)$);
	\node[above=5pt of a,anchor=base] {$n, n$};
	\node[above=5pt of b,anchor=base] {$n, s$};
	\node[above=5pt of c,anchor=base] {$s, n$};
	\node[above=5pt of d,anchor=base] {$s, s$};

	\coordinate (e) at ($(m.north west)!0.125!(m.south west)$);
	\coordinate (f) at ($(m.north west)!0.375!(m.south west)$);
	\coordinate (g) at ($(m.north west)!0.625!(m.south west)$);
	\coordinate (h) at ($(m.north west)!0.875!(m.south west)$);
	\node[left=1pt of e,text width=0.62cm]  {$n, n$};
	\node[left=1pt of f,text width=0.62cm]  {$n, s$};
	\node[left=1pt of g,text width=0.62cm]  {$s, n$};
	\node[left=1pt of h,text width=0.62cm]  {$s, s$};

	\node[above=14pt of m.north] (species Y) {Species $Y$};
	\node[left=1.2cm of m.west,rotate=90,align=center,anchor=center] {Species $X$};

	\end{tikzpicture}

	\vspace{0.3cm}

	We note $s$ as the action ``share information'' and $n$ as the action ``do not share information''. Then, for instance, $(n, s)$, is
	a short expression of ``\textbf{n}ot sharing in the first time-step, and \textbf{s}haring in the second time-step''. Let $v, w$ be strategies
	in $\{(n,n), (n,s), (s,n), (s,s)\}$. The payoff of species $X$ when species $X$ plays strategy $v$ and species $Y$ plays strategy $w$
	is represented by $u_x(v, w)$.

	\vspace{0.3cm}

	We say a strategy $v^\ast \in \{(n,n), (n,s), (s,n), (s,s)\}$ is \textbf{strictly dominant} if \\
	\begin{equation}
	\forall v^\prime \in \{(n,n),(n,s),(s,n),(n,n)\}, v^\prime \neq v^\ast, \text{ we have that } u_x(v^\ast, w) > u_x(v^\prime, w)
	\end{equation}

	\vspace{0.1cm}

	We say a strategy $v^\ast \in \{(n,n), (n,s), (s,n), (s,s)\}$ is \textbf{weakly dominant} if \\
	\begin{equation}
	\forall v^\prime \in \{(n,n),(n,s),(s,n),(n,n)\}, v^\prime \neq v^\ast, \text{ we have that } u_x(v^\ast, w) \geq u_x(v^\prime, w)
	\end{equation}
	with at least one strategy $v^\prime$ giving a strict inequality.

	}
  \vskip10pt
 }
}

\vspace{1cm}

\begin{figure*}[ht]
	\centering
	\begin{minipage}[b]{.33\textwidth}
		\centering
		\includegraphics[scale=0.9]{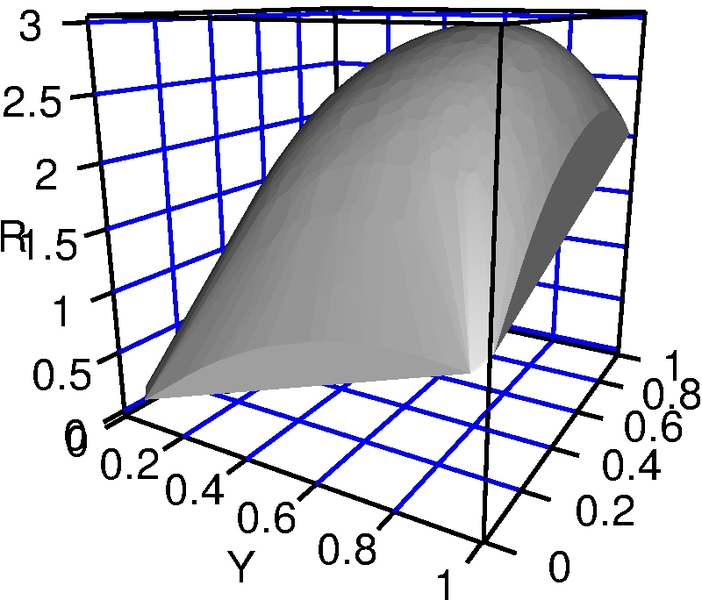}
		\subcaption{No dominant \\ strategies}
		\label{fig:nds}
	\end{minipage}\hfill
	\begin{minipage}[b]{.33\textwidth}
		\centering
		\includegraphics[scale=0.9]{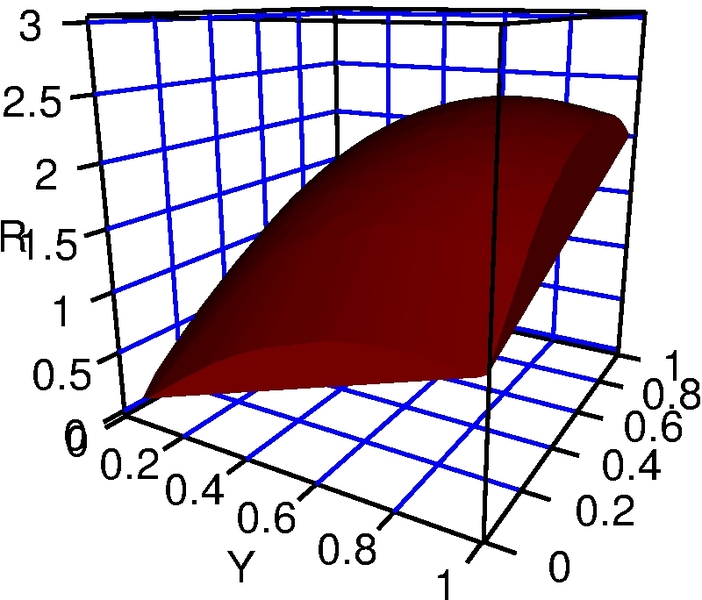}
		\subcaption{Not sharing weakly dominant}
		\label{fig:nwd}
	\end{minipage}\hfill
	\begin{minipage}[b]{.34\textwidth}
		\centering
		\includegraphics[scale=0.9]{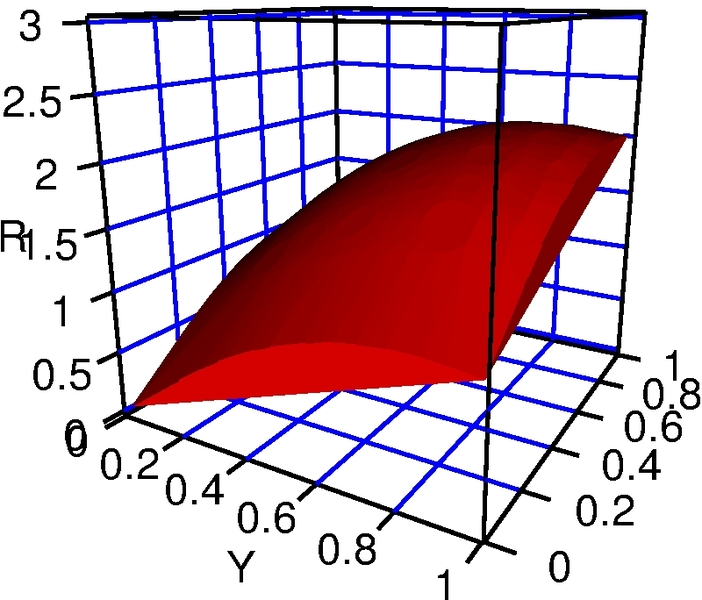}
		\subcaption{Not sharing strictly dominant}
		\label{fig:nsd}
	\end{minipage}\hfill
	\caption{(see in colour) We analyse the payoff matrix resulting from initial values in $\mathcal{X}\times\mathcal{Y}\times\mathcal{R}$.
		We obtained five non-overlapped volumes classifying the strategies: 
		(a) The grey volume corresponds to initial values where the optimal strategy of species $X$
		is conditioned on the strategy of species $Y$, and therefore there are no dominant strategies. For all points above this volume,
        sharing information is weakly dominant. (b) The dark-red volume corresponds to initial values where not sharing information is weakly
        dominant for species $X$. (c) The red volume corresponds to initial
		values where not sharing information is strictly dominant for species $X$. Finally, for all points below volume (c), species $X$
		will get extinct independently of its behaviour.}
	\label{fig:domplots}
\end{figure*}

\begin{center}
\begin{figure}[ht]
	\centering
	\noindent\begin{minipage}[b]{.05\textwidth}
		\rotatebox{90}{\tiny{~~~~~~~~~~$X$ population density}}
	\end{minipage}%
	\noindent\begin{minipage}[b]{.19\textwidth}
		\includegraphics[scale=0.6]{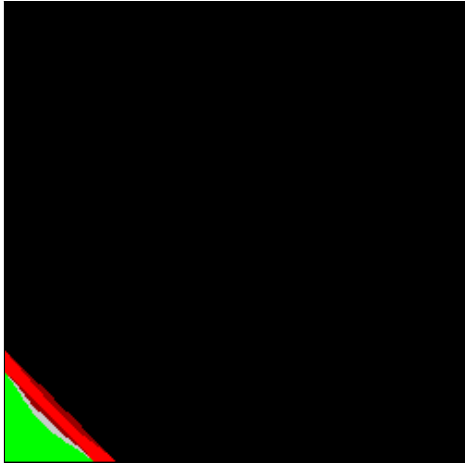}
		\vspace*{0.1cm}\hspace*{0.3cm}{\tiny{$Y$ population density}}
		\subcaption{$R = 0.4$}
		\label{fig:r0.4}
	\end{minipage}%
	\noindent\begin{minipage}[b]{.19\textwidth}
		\includegraphics[scale=0.6]{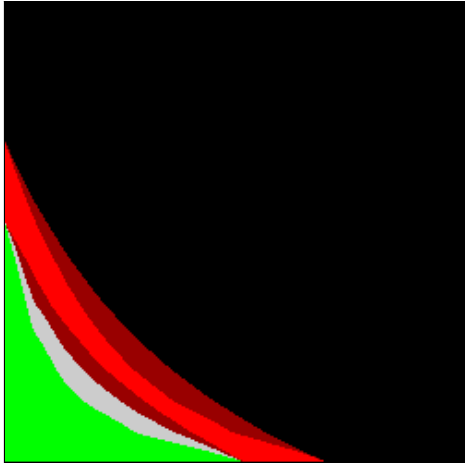}
		\vspace*{0.1cm}\hspace*{0.3cm}{\tiny{$Y$ population density}}
		\subcaption{$R = 0.9$}
		\label{fig:r0.9}
	\end{minipage}%
	\noindent\begin{minipage}[b]{.19\textwidth}
		\includegraphics[scale=0.6]{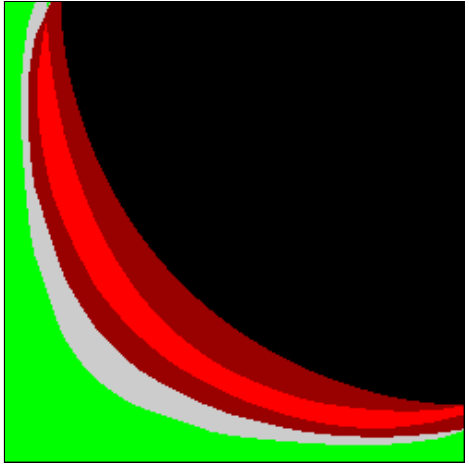}
		\vspace*{0.1cm}\hspace*{0.3cm}{\tiny{$Y$ population density}}
		\subcaption{$R = 1.2$}
		\label{fig:r1.2}
	\end{minipage}%
	\noindent\begin{minipage}[b]{.19\textwidth}
		\includegraphics[scale=0.6]{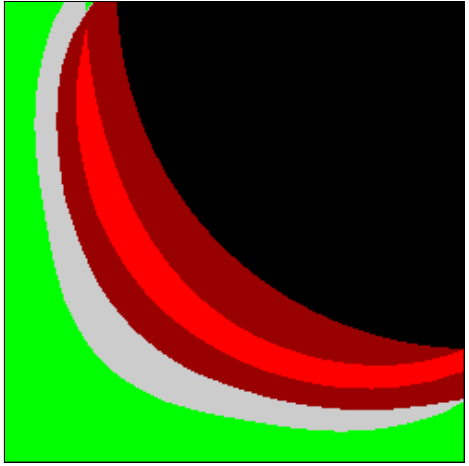}
		\vspace*{0.1cm}\hspace*{0.3cm}{\tiny{$Y$ population density}}
		\subcaption{$R = 1.4$}
		\label{fig:r1.4}
	\end{minipage}%
	\noindent\begin{minipage}[b]{.19\textwidth}
		\includegraphics[scale=0.6]{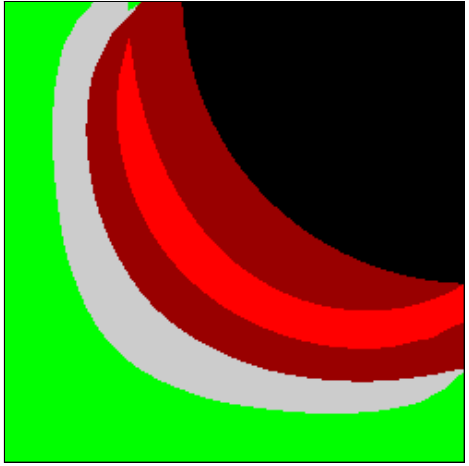}
		\vspace*{0.1cm}\hspace*{0.3cm}{\tiny{$Y$ population density}}
		\subcaption{$R = 1.6$}
		\label{fig:r1.6}
	\end{minipage}%
	\vspace{0.35cm}
	\centering
	\noindent\begin{minipage}[b]{.05\textwidth}
		\rotatebox{90}{\tiny{~~~~~~~~~~$X$ population density}}
	\end{minipage}%
	\begin{minipage}[b]{.19\textwidth}
		\includegraphics[scale=0.6]{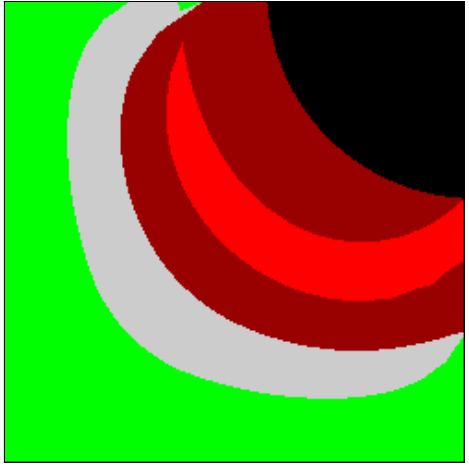}
		\vspace*{0.1cm}\hspace*{0.3cm}{\tiny{$Y$ population density}}
		\subcaption{$R = 1.8$}
		\label{fig:r1.8}
	\end{minipage}%
	\begin{minipage}[b]{.19\textwidth}
		\includegraphics[scale=0.6]{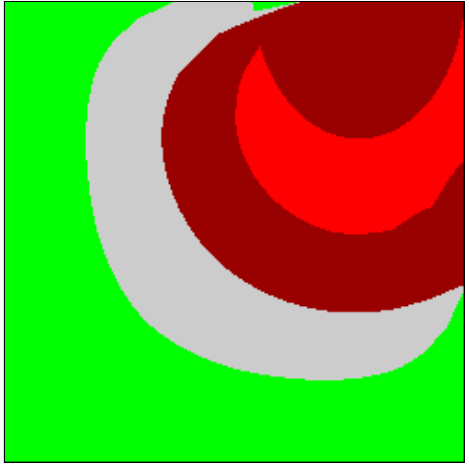}
		\vspace*{0.1cm}\hspace*{0.3cm}{\tiny{$Y$ population density}}
		\subcaption{$R = 2.0$}
		\label{fig:r2.0}
	\end{minipage}%
	\begin{minipage}[b]{.19\textwidth}
		\includegraphics[scale=0.6]{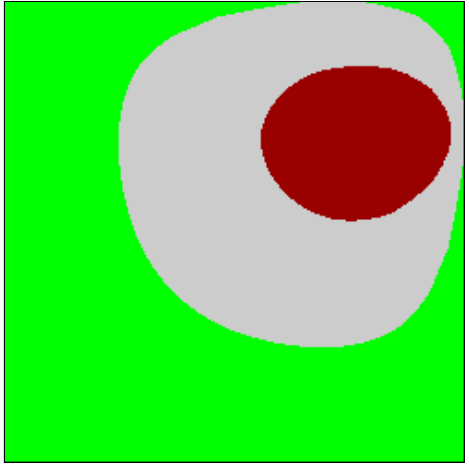}
		\vspace*{0.1cm}\hspace*{0.3cm}{\tiny{$Y$ population density}}
		\subcaption{$R = 2.3$}
		\label{fig:r2.3}
	\end{minipage}%
	\begin{minipage}[b]{.19\textwidth}
		\includegraphics[scale=0.6]{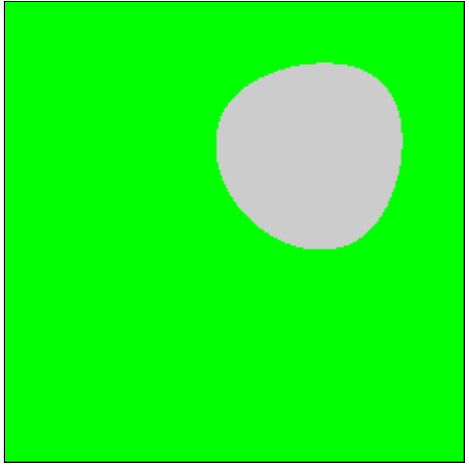}
		\vspace*{0.1cm}\hspace*{0.3cm}{\tiny{$Y$ population density}}
		\subcaption{$R = 2.8$}
		\label{fig:r2.8}
	\end{minipage}%
	\begin{minipage}[b]{.19\textwidth}
		\includegraphics[scale=0.6]{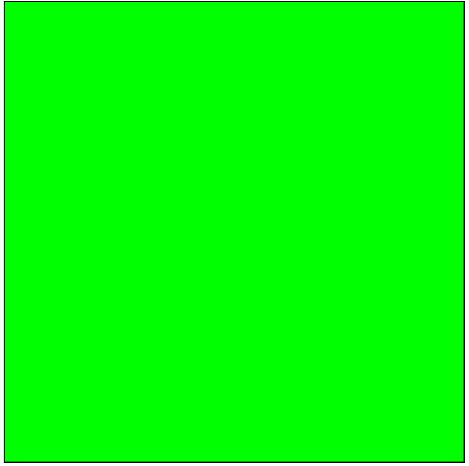}
		\vspace*{0.1cm}\hspace*{0.3cm}{\tiny{$Y$ population density}}
		\subcaption{$R = 3.0$}
		\label{fig:r3.0}
	\end{minipage}%
	\caption{(see in colour) We analyse the payoff matrix from $250^2$ initial values of population densities of species $X$ and
			$Y$ uniformly distributed in $\mathcal{X}\times\mathcal{Y}$, with the amount of resources $R$ fixed. Green points corresponds
			to sharing information being a weakly dominant strategy for species $X$. Grey points corresponds to initial values where there
			is no dominant strategy for species $X$. Red points represent values for which not sharing information is strictly
			dominant for species $X$. Dark-red points corresponds to values where not sharing information is weakly dominant for
			species $X$. Finally, black points represent values where species $X$ goes extinct regardless of its strategy.}
	\label{fig:gameanalysis}
\end{figure}
\end{center}

In Fig. \ref{fig:domplots}, we show a classification of the initial values of $R$, $X$ and $Y$ based on the resulting payoff matrices.
The plots shown are the result of computing the convex hull on the classified points, and for this reason the ``bottom'' part of the
volumes appear to be straight. This will hopefully become clear by looking at Fig. \ref{fig:gameanalysis}.

Figure \ref{fig:nds} shows situations where there is no dominant strategy --- the optimal one is conditioned on the other species' strategy.
For instance, in Table \ref{tab:nds} we show an example of such payoff matrix. Here, the payoff of species $X$ when it
plays $(n,s)$ (short for ``\textbf{n}ot sharing in the first time-step, and \textbf{s}haring in the second time-step'') and species $Y$ plays
$(s,s)$ is higher than when species $X$ plays $(n,n)$ while keeping $Y$'s strategy the same. While this may seem counter-intuitive, since the
returns (in environmental information) for species $X$ when sharing information in the second time-step are not perceived by it due to the locality
of the look-ahead, it nevertheless increases its payoff since $Y$'s population mortality is increased. In Sec. \ref{sec:discussion} we discuss
how such situations could be analysed. 

For initial values where the amount of resources is higher than those of the volume of Fig. \ref{fig:nds}, sharing information is weakly
dominant for species $X$. This situation corresponds to amounts of resources such that the consumption of both populations after two time-steps
does not deplete them, and hence sharing information cannot hurt a species, since its growth will not be affected. Moreover, sharing information
would be beneficial, in cases where the other species shares back.

\begin{table}[H]
	\centering
	\begin{tikzpicture}
	
	\matrix[matrix of math nodes,every node/.style={align=center,text width=2.5cm},row sep=0.08cm,column sep=0.50cm, nodes in empty cells] (m)
	{
	-0.35204577&-0.22381541&-0.20745971&-0.11033376\\
	-0.35204577&-0.22381541&-0.16294896&-0.09836764\\
	-0.35204577&-0.22381541&-0.20745971&-0.11033376\\
	-0.35204577&-0.23964398&-0.16294896&-0.15442442\\
	};
	
	\draw (m.north east) rectangle (m.south west);
	\draw (m.north) -- (m.south);
	\draw (m.east) -- (m.west);
	
	\draw ($(m.north west)!0.25!(m.north east)$) -- ($(m.south west)!0.25!(m.south east)$);
	\draw ($(m.north west)!0.75!(m.north east)$) -- ($(m.south west)!0.75!(m.south east)$);
	
	\draw ($(m.north west)!0.25!(m.south west)$) -- ($(m.north east)!0.25!(m.south east)$);
	\draw ($(m.north west)!0.75!(m.south west)$) -- ($(m.north east)!0.75!(m.south east)$);
	
	\coordinate (a) at ($(m.north west)!0.125!(m.north east)$);
	\coordinate (b) at ($(m.north west)!0.375!(m.north east)$);
	\coordinate (c) at ($(m.north west)!0.625!(m.north east)$);
	\coordinate (d) at ($(m.north west)!0.875!(m.north east)$);
	\node[above=5pt of a,anchor=base] {$n, n$};
	\node[above=5pt of b,anchor=base] {$n, s$};
	\node[above=5pt of c,anchor=base] {$s, n$};
	\node[above=5pt of d,anchor=base] {$s, s$};
	
	\coordinate (e) at ($(m.north west)!0.125!(m.south west)$);
	\coordinate (f) at ($(m.north west)!0.375!(m.south west)$);
	\coordinate (g) at ($(m.north west)!0.625!(m.south west)$);
	\coordinate (h) at ($(m.north west)!0.875!(m.south west)$);
	\node[left=2pt of e,text width=0.62cm]  {$n, n$};
	\node[left=2pt of f,text width=0.62cm]  {$n, s$};
	\node[left=2pt of g,text width=0.62cm]  {$s, n$};
	\node[left=2pt of h,text width=0.62cm]  {$s, s$};
	
	\node[above=14pt of m.north] (species Y) {Species $Y$};
	\node[left=1.4cm of m.west,rotate=90,align=center,anchor=center] {Species $X$};
	
	\end{tikzpicture}
	\caption{Example of a payoff matrix where there is no dominant strategy for species $X$. The initial values for this specific matrix where
			$X = 0.5$, $Y = 0.2$ and $R = 1.8$. Each cell contains the growth rate of species $X$ when each species plays the correspondent sequence of actions.
			}
	\label{tab:nds}
\end{table}

For cases where the potential benefits of having extra information from the other species is always outweighed by the decrease in the populations
due to the diminished resources, then not sharing information is a strictly dominant strategy (see Fig. \ref{fig:nsd}). In Fig. \ref{fig:nwd} we
show the volume corresponding to initial values where not sharing is weakly dominant. This volumes ``encloses'' the one shown in Fig. \ref{fig:nsd},
where initial values can be distinguished within two types: in the first one, resources are sufficient for both species to share information only
in the first time-step, and therefore, species $X$ achieves the same payoff playing either $(n,n)$ or $(s,n)$ when $Y$ plays either $(n,n)$ or $(s,n)$
(see Table \ref{tab:nwd_1}). In these situations, the reduction of resources after the first time-step makes sharing information damaging as the
subsequent action.

\begin{table}[htp]
	\centering
	\begin{tikzpicture}
	
	\matrix[matrix of math nodes,every node/.style={align=center,text width=2.5cm},row sep=0.08cm,column sep=0.50cm, nodes in empty cells] (m)
	{
	-0.46579296&-0.31965691&-0.27855031&-0.25980521\\
	-0.48390350&-0.63747731&-0.34778310&-0.59688452\\
	-0.46579296&-0.40127033&-0.27855031&-0.33887903\\
	-0.59319779&-0.79195649&-0.44195076&-0.64172489\\
	};
	
	\draw (m.north east) rectangle (m.south west);
	\draw (m.north) -- (m.south);
	\draw (m.east) -- (m.west);
	
	\draw ($(m.north west)!0.25!(m.north east)$) -- ($(m.south west)!0.25!(m.south east)$);
	\draw ($(m.north west)!0.75!(m.north east)$) -- ($(m.south west)!0.75!(m.south east)$);
	
	\draw ($(m.north west)!0.25!(m.south west)$) -- ($(m.north east)!0.25!(m.south east)$);
	\draw ($(m.north west)!0.75!(m.south west)$) -- ($(m.north east)!0.75!(m.south east)$);
	
	\coordinate (a) at ($(m.north west)!0.125!(m.north east)$);
	\coordinate (b) at ($(m.north west)!0.375!(m.north east)$);
	\coordinate (c) at ($(m.north west)!0.625!(m.north east)$);
	\coordinate (d) at ($(m.north west)!0.875!(m.north east)$);
	\node[above=5pt of a,anchor=base] {$n, n$};
	\node[above=5pt of b,anchor=base] {$n, s$};
	\node[above=5pt of c,anchor=base] {$s, n$};
	\node[above=5pt of d,anchor=base] {$s, s$};
	
	\coordinate (e) at ($(m.north west)!0.125!(m.south west)$);
	\coordinate (f) at ($(m.north west)!0.375!(m.south west)$);
	\coordinate (g) at ($(m.north west)!0.625!(m.south west)$);
	\coordinate (h) at ($(m.north west)!0.875!(m.south west)$);
	\node[left=2pt of e,text width=0.62cm]  {$n, n$};
	\node[left=2pt of f,text width=0.62cm]  {$n, s$};
	\node[left=2pt of g,text width=0.62cm]  {$s, n$};
	\node[left=2pt of h,text width=0.62cm]  {$s, s$};
	
	\node[above=14pt of m.north] (species Y) {Species $Y$};
	\node[left=1.4cm of m.west,rotate=90,align=center,anchor=center] {Species $X$};
	
	\end{tikzpicture}
	\caption{Typical payoff matrix where not sharing information is weakly dominant for species $X$. The initial values for this specific
			matrix where $X = 0.28$, $Y = 0.76$ and $R = 1.8$. Each cell contains the growth rate of species $X$ when each species plays the correspondent sequence of actions.
			}
	\label{tab:nwd_1}
\end{table}

For the second type of initial values, sharing information in the second time-step causes complete depletion of resources and therefore species'
$X$ subsequent extinction. A typical example of the payoff matrix for these cases is shown in Table \ref{tab:nwd_2}, where we can see why the
strategy $(n,n)$ is not strictly dominant: if a species shares information in the second time-step, then species $X$ will get extinct no matter
what the other species does, obtaining the same payoff for all the other species' options. Let us note here that a growth rate of $-1.0$ implies
the extinction of the species, since this value is a lower bound for the growth rate, and can only be achieved when the proportion of the
population that acquired environmental information is zero (which means that both populations completely die out). Finally, for all initial values
of $X$, $Y$ and $R$ below the volume shown in Fig. \ref{fig:nwd}, species will go extinct independently of their behaviour.

\begin{table}[htp]
	\centering
	\begin{tikzpicture}
	
	\matrix[matrix of math nodes,every node/.style={align=center,text width=2.5cm},row sep=0.08cm,column sep=0.50cm, nodes in empty cells] (m)
	{
	-0.59296608&-1.00000000&-0.52533449&-1.00000000\\
	-1.00000000&-1.00000000&-1.00000000&-1.00000000\\	
	-0.65597890&-1.00000000&-0.59296600&-1.00000000\\	
	-1.00000000&-1.00000000&-1.00000000&-1.00000000\\	
	};
	
	\draw (m.north east) rectangle (m.south west);
	\draw (m.north) -- (m.south);
	\draw (m.east) -- (m.west);
	
	\draw ($(m.north west)!0.25!(m.north east)$) -- ($(m.south west)!0.25!(m.south east)$);
	\draw ($(m.north west)!0.75!(m.north east)$) -- ($(m.south west)!0.75!(m.south east)$);
	
	\draw ($(m.north west)!0.25!(m.south west)$) -- ($(m.north east)!0.25!(m.south east)$);
	\draw ($(m.north west)!0.75!(m.south west)$) -- ($(m.north east)!0.75!(m.south east)$);
	
	\coordinate (a) at ($(m.north west)!0.125!(m.north east)$);
	\coordinate (b) at ($(m.north west)!0.375!(m.north east)$);
	\coordinate (c) at ($(m.north west)!0.625!(m.north east)$);
	\coordinate (d) at ($(m.north west)!0.875!(m.north east)$);
	\node[above=5pt of a,anchor=base] {$n, n$};
	\node[above=5pt of b,anchor=base] {$n, s$};
	\node[above=5pt of c,anchor=base] {$s, n$};
	\node[above=5pt of d,anchor=base] {$s, s$};
	
	\coordinate (e) at ($(m.north west)!0.125!(m.south west)$);
	\coordinate (f) at ($(m.north west)!0.375!(m.south west)$);
	\coordinate (g) at ($(m.north west)!0.625!(m.south west)$);
	\coordinate (h) at ($(m.north west)!0.875!(m.south west)$);
	\node[left=2pt of e,text width=0.62cm]  {$n, n$};
	\node[left=2pt of f,text width=0.62cm]  {$n, s$};
	\node[left=2pt of g,text width=0.62cm]  {$s, n$};
	\node[left=2pt of h,text width=0.62cm]  {$s, s$};
	
	\node[above=14pt of m.north] (species Y) {Species $Y$};
	\node[left=1.4cm of m.west,rotate=90,align=center,anchor=center] {Species $X$};
	
	\end{tikzpicture}
	\caption{Typical payoff matrix where not sharing information is weakly dominant for species $X$. The initial values for this specific
			matrix where $X = 0.6$, $Y = 0.6$ and $R = 1.8$. Each cell contains the growth rate of species $X$ when each species plays the correspondent sequence of actions.
			}
	\label{tab:nwd_2}
\end{table}

As a complement, in Fig. \ref{fig:gameanalysis} we show 2D plots by fixing the amount of resources to $10$ different values, in each of these
values analysing $250^2$ points uniformly distributed in $\mathcal{X}\times\mathcal{Y}$. The black zone corresponds to initial values of population
densities in which species $X$ go extinct regardless of its strategy. In these plots, we can more clearly visualise the described ``enclosure'' of
the volume shown in Fig. \ref{fig:nsd}. Additionally, we can better appreciate the relationship between resources and population densities. Relative
terms used to describe the amounts of resources such as ``scarce'' of ``abundant'' are directly correlated with the areas shown in the plots.
For instance, scarce resources are in correspondence with red, dark-red and black areas; while abundant resources are in correspondence with
green areas.

As resources increase in absolute values, the range of population densities for which resources are scarce gets smaller. In Fig. \ref{fig:r1.8},
for instance, we see the red area surrounded by the dark-red areas, where values of $X$ between the red and grey areas are those of the type
exemplified by the payoff matrix shown in Table \ref{tab:nwd_1}; and those between the red and the black areas are those of the type exemplified by the
payoff matrix shown in Table \ref{tab:nwd_2}.

The scarcity area disappears approximately when $R \geq 2.4$. Opposed to this scenario, when resources decrease, the range of population
densities for which resources are abundant also gets reduced, corresponding only to small population densities (see Fig. \ref{fig:r0.4} and
\ref{fig:r0.9} for examples). In grey areas, resources can be considered neither scarce nor abundant. This area presents an incentive for
species to coordinate behaviour, as we discuss in Sec. \ref{sec:analysisnd}. For resources values approximately of $R > 2.8$, then
sharing information is always weakly dominant.


\section{Discussion}
\label{sec:discussion}

Our theoretical model integrates two behaviours of bacteria, bet-hedging and cell-to-cell communication, which are usually studied
in isolation \cite{Perkins2009}. Bacterial cells follow a bet-hedging strategy, incorporating density-dependent environmental
information into their decision-making process. Therefore, a cell's communication behaviour influences the long-term growth rate of
other cells. Assuming within-species communication, what can we say about the dynamics of information exchange between species?


As we have seen, environmental information is translated into long-term growth rate. Thus, cells that acquire environmental information
will have an advantage over those that do not. While acquiring information (other than that which a species already has) depends solely
on whether the other species shares information, the potential recipient species can actively increase the amount of information the other
species may provide in the future, by sharing information with it. The cost of this investment is zero when resources are sufficient for
the consumption of both populations. Therefore, when resources
are abundant, cooperative strategies between species of bacterial cells will out-compete those where none or only one of the species
cooperate.

When the consumption of resources by the populations results in a reduction of resources, then the cost of sharing information is not
zero any more, and it is related to the loss in the species' environmental information caused by the diminished proportion of cells
perceiving the environment. In cases where a species does not share back information, the other species will always lose its investment
in the first species' growth. On the other hand, when the first species does share in return, and depending on initial values, it may pay
off to invest. These situations correspond to the volume shown in Fig. \ref{fig:nds}, where there are no dominant strategies. 

As resources become scarcer, the cost of sharing information becomes higher, and eventually the losses caused by the other species having
extra information outweigh any possible benefit (volumes in Fig. \ref{fig:nwd} and Fig. \ref{fig:nsd}). In this scenario, bacterial cells
developing an antagonistic behaviour will out-compete those that do not. 

It becomes clear from our analysis that there is an indirect cost for sharing information which is relative to the amount of resources
and to the population densities. As noted in \cite{Lachmann2000}, the immediate cost of sharing information is different from that of
sharing a resource (such as food) (although there is evidence that there is a cost associated with signal production in bacteria
\cite{Keller2006, Diggle2007}, incorporating this feature in the model would not qualitatively change the results). Namely, in the latter,
the shared amount equals the losses of an organism and the gains of another, while in the former, as well as in our model, sharing
information does not incur any immediate cost. However, the indirect cost of sharing information is given by the decrease in a
species' environmental information, which is zero in abundance, and increases as resources become scarcer. Our model captures the relative
value of resources, which dominates the species' communication behaviour.


The transition in the dominant strategy, from cooperation to antagonism, results from a change in the availability of resources: in
abundance, cooperative species out-compete non-cooperative ones, while in scarcity, antagonistic species out-compete non-antagonistic
ones. This is supported by the results obtained in \cite{Requejo2011, Requejo2012}, where there is also a transition 
from cooperation to defection depending on available resources. However, in this work, cooperation results dominant in scarcity of
resources, while defection is dominant in abundance. Although the results seem contradictory, the difference comes from the assumptions:
for unlimited resources, cooperation, in our model, gives players an advantage, while, in their model, defection does. Therefore,
qualitatively, results support each other: in our model, the game is equivalent to a Prisoner's Dilemma in scarcity of resources and to
the Harmony Game in abundance; while in their resource model, the relationship is the opposite. 

Finally, a comment regarding the complexitiy of the computation for bacteria of the communication strategy: contexts (initial values) where the species 
does not share information belong to a well-defined region that can be approximated using a threshold value. The same is valid for
contexts where the species does share information. For contexts with no dominant strategies, more complicated computations are needed.
One prediction that could be possible would be that simple organisms would either avoid this area because it requires more complex
computation, or indeed, that even very simple organisms that operate in this region do have more complex decision-making cascades.
However, we are cautious making a concrete numerical prediction, because for an experimental test a more precise understanding of the
dynamics will be necessary.

\subsection{Possible analysis of the no-dominance volume}
\label{sec:analysisnd}

For situations where there is no dominant strategy, it would still be desirable to be able to predict the communication behaviour of the
species. In this transition, species would benefit from predicting the other species' behaviour, such that they can adjust their response
accordingly. This leads to two types of games: Snow Drift, where players benefit by playing the opposite of the co-player; and Stag Hunt,
where players benefit by coordinating their actions. Or, it can result in a more intricate game, where a species would share as much
information as it can as long as it does not deplete the resources. In this case, the expected outcome of an evolutionary process may be
a mixed population of cells sharing information and cells not sharing information; or we could have the species, at the population level,
choosing how much information to share with the other.

The latter could be analysed by allowing the species' population to share partial information, and the Nash equilibria would indicate the
possible outcomes. Here, it may be interesting to consider if playing mixed strategies is equivalent (in utility, growth rate in
our case) to playing a linear mix of pure strategies, as it follows from any rational agent satisfying the axioms as defined in
\cite{Neumann1947}. For instance, if playing a linear mix of pure strategies (\emph{i.e.} sharing partial information) achieves a higher
utility than playing mixed strategies, then this may be indicative of the outcome of an evolutionary process where there are no dominant
strategies.

For the former case, we allow each cell to take two actions, sharing or not sharing their information with the other species. This translates
into $n + 1$ actions at the population level, since cells are clones of each other, and it is sufficient to count how many cells share information
($n$ is the number of cells in the population). 

\subsection{Modelling bet-hedging mechanisms}
\label{sec:bh_mechanisms}

In the presented model, we made a strong assumption in relation to the interpretation of the information a species obtains. Namely, we
assumed that all the information communicated by one species was unambiguously interpreted by the other species, and vice versa, and they
were both able to translate this information into the optimal bet-hedging strategy. However, in biological systems, information can be,
for instance, ambiguous, meaningless or false, leading to the implementation of sub-optimal bet-hedging strategies.

The incorporation of a bet-hedging mechanism into the model would require explicitly modelling the actions of cells, where an
action is developing into a particular phenotype. The policy of a cell would indicate how it translates the perceived information into
actions. Now, in order for cells to be able to communicate, one of the following properties need to hold: either the identity of the sender is
known, in which case the transmitted ``message'' can be fully interpreted (further assuming absence of noise in the used channel); or
they would need to agree on a common language: that is, they would have identical (or similar) policies for interpreting messages, such that,
no matter who the sender is, the information can still be interpreted \cite{Burgos2014, Burgos2015}. In other words, where identification
of the sender is not possible, then a common language is necessary in order to make sense of the information. Such framework would allow further
interesting dynamics, such as parasitism, where some cells convey ``false'' information for the detriment of other cell's predictions
\cite{Burgos2015}. We believe that these are essential aspects to include in the study of bet-hedging mechanisms.

\subsection{Other interpretations of the model}
\label{sec:innovations}

Although we presented the model in a biological context, it could as well be considered in other contexts, such as economics. For instance,
we could think of two software companies sharing the same market with the option to adopt two different models: open source or closed
source software production. Assuming a high demand for such products, a free flow of innovations would allow higher growth rates (in terms
of returns) for both companies, while, when competing for demand, a closed source model would benefit both of them.

Particularly in our model, we could consider innovations to be environmental information which is not already present in the
collective information of a population. Then, if one company is more proficient than another company in developing software for a
particular niche, the latter could benefit from the innovations of the former to expand its market (in our model, we assume each species
is more proficient in capturing different aspects of the environment). Then, acquired innovations would be translated into higher growth
rates.

It is important to note that, in our model (and under this consideration), innovations are implicitly assumed to increase with population
size (see Fig. \ref{fig:envinfo}). A more truthful model should distinguish the information that is incorporated into a population (which
could be redundant, innovative, or of other types), as well as how the new information is integrated with the existing information (whether
it is compatible or not). Finally, our model assumes that the knowledge of \emph{how} to perform the actions necessary to survive for
certain conditions (develop a certain phenotype for bacteria, or, for a software company, produce a particular code) is available for both
species or companies, and thus innovations here should be understood as new knowledge which improves the prediction of future conditions.

To consider other types of innovations, such as those that would allow the optimisation of the processes producing the actions, or even
innovations that would result in new actions enabling expansion, a more comprehensive and complex model would be needed. These types of
innovations allow bacteria, for example, to incorporate traits through lateral gene transfer such as antibiotic resistance, virulence
attributes and metabolic properties \cite{Ochman2000}. In the same way, software companies can integrate efficient modules performing
specific tasks into their projects. As stated above, the incorporation of foreign information raises issues of redundancy and language
compatibility, where reading a gene or executing a module would have to be possible, and the results of such actions would have to be
integrated with the rest of the system. 

In relation to this, we could also interpret our model in the framework of cellular evolution, where there is a transition from horizontal
exchange of genetic material between primitive cells (cooperation) to a stage dominated by vertical transfer (antagonism). In early stages
of evolution, primitive cells would constantly exchange genetic material through horizontal gene transfer (HGT) \cite{Woese2002,
Woese2004}.These can be considered ``innovations'', and would allow them to achieve higher growth rates. However, this would also present
the problem we have just discussed about incorporating foreign information to a functioning system. This problem was considered in
\cite{Vetsigian2006}, where they model the evolution of the genetic code accounting for universality and optimality. In their work, they
consider the genetic code ``not only as a protocol for encoding amino acid sequences in the genome, but also an innovation-sharing
protocol'' \cite{Vetsigian2006}. While our model ignores the intricate aspects of exchange of genetic material, it offers a high level
interpretation of the transition from HGT to vertical gene transfer (VGT). Other studies investigated the evolution of the genetic code in 
this context, serving also as inspiration for this work \cite{Piraveenan2007, Polani2008, Obst2011}.

\subsection{Stigmergy}
\label{sec:stigmergy}

As recently noted in a study of self-organisation in bacterial biofilms \cite{Gloag2013b}, bacterial communication can be considered as
a type of stigmergy \cite{Grasse1959}, where cells modify their environment by releasing chemical signals and influence the behaviour of
the cells perceiving them. This results in a coordinated collective behaviour without the necessity of a central control. In the mentioned
study, the expansion in biofilms of the bacterium \emph{Pseudomonas aeruginosa} is analysed. This bacterium has the ability to remodel
its substratum to form an interconnected network of trails, which guides the transit of cells, and uses extracellular DNA to facilitate
traffic flow through it \cite{Gloag2013a, Gloag2013b}.

Many distinctions have been made on the concept of stigmergy, such as sematectonic or marker-based \cite{Wilson1975},
quantitative or qualitative \cite{Theraulaz1999}. These distinctions can be considered orthogonal \cite{Parunak2006}, and they are
important to describe in more precision the system in question: for instance, quorum sensing can be considered marker-based and
quantitative, but also qualitative (bacteria recognises different chemical signals, for example in cross-species talking). Other
distinctions have been proposed, one related to the duration of modifications, transient or persistent, and the other related to the
structure of the population, termed broadcast or narrowcast \cite{Heylighen2011}. 

Specifically in this study, we do not explicitly model the mentioned aspects of stigmergy, but by considering our model in the framework
of stigmergy, they contribute to more accurately describe the assumptions made. First, the communication between cells is assumed to be
instantaneous and transient, since in every time-step the previously shared information is not considered. Second, information is
broadcasted, since every individual cell perceives the output of every other cell. Third, information is qualitative, as shown in Fig.
\ref{fig:envinfo}, where the information of a population increases with population size. Finally, whether communication is marker-based
or sematectonic, nothing particular is assumed in the model.


\subsection{Multilevel selection}
\label{sec:multilevelselection}

As mentioned in the introduction, we assumed in our model within-species communication in order to simplify the game-theoretic analysis.
However, it would be desirable to analyse whether individual cells would share information with other cells of the same species or not by
considering a communication strategy for each cell. 

However, in this scenario, natural selection would operate at multiple levels \cite{Michod1999, Keller1999}, and a preference between
sharing information with the same species rather than with other species may need to be considered. In our particular setting, species
capture different aspects of the environment on which they depend, and we can speculate that the preference, at least initially, would be
towards the other species, who contributes more to the total environmental information (see Sec. \ref{sec:betweencomm}).

This seems contrary to kin selection \cite{Hamilton1964, MaynardSmith1964}, where individuals would prefer to cooperate with individuals of
the same species (and thus maximising inclusive fitness). Instead, because of our assumption of global competition on resources, there is
as much competition between kins as there is between non-kins, and since the other species provides more information about the environment,
interactions with members of the other species would be preferred. Had we assumed that the contribution in environmental information
from members of the same species was larger than that of the other, then the preference of cooperation would be toward kins. For the
latter situation, such behaviours have been observed in the pathogen \emph{Pseudomonas aeruginosa} \cite{Diggle2007}. Moreover, the same
bacterium diminishes kin cooperation as the scale of competition becomes more local \cite{Griffin2004}. The scale at which species compete
would have a significant effect in the communication behaviour of individual cells \cite{Griffin2004, Platt2009}. 


\section{Conclusion}

To conclude, we presented an information-theoretic model which integrates two aspects of bacterial behaviour, bet-hedging and cell-to-cell
communication. While simple, several important aspects of communication were captured by our model: we related the communication behaviour
of species to the relative availability of resources, which can be classified into three main regimes. Species cooperating in abundance of
resources would benefit, while they would behave antagonistically in scarcity. In this transition, for the situations in-between, species
would have an incentive to coordinate their behaviours, adapting in response to each other's strategies.



\appendix

\section{Information theory}
\label{sec:inftheory}

In this section we briefly introduce some basic concepts of Information theory. For an in-depth treatment, we refer the reader to
\cite{Cover2002}. A measure for the uncertainty of a random variable is given by the \emph{entropy} of that random variable. For
instance, the entropy of the environment $E$ for a set of environmental conditions $\mathcal{E}$ is defined as

\begin{equation}
H(E) = -\sum\limits_{e\;\in\;\mathcal{E}} p(e)~\log_2~p(e)
\end{equation}

The \emph{conditional entropy} measures the uncertainty of a random variable once another random variable has been observed.
For instance, the uncertainty of the environment once an individual $i$ of species $X$ has acquired information from its sensors
is defined as

\begin{equation}
H\left(E \;\middle\vert\; S_x\right) = -\sum\limits_{s_x} p(s_x) \sum\limits_{e} p\left(e \;\middle\vert\; s_x\right)~\log_2~p\left(e \;\middle\vert\; s_x\right)
\end{equation}

The \emph{mutual information} measures the reduction in the uncertainty of a random variable due to the knowledge of
another random variable. For instance, the reduction in the uncertainty of the environment for an individual $i$ of species $X$
due to the knowledge of its acquired sensory information is defined as

\begin{equation}
I(E \;;\; S_x) = H(E) - H\left(E \;\middle\vert\; S_x\right) = \sum\limits_{e, s_x} p(e, s_x)~\log_2~\frac{p(e, s_x)}{p(e)p(s_x)}
\end{equation}

Finally, the \emph{Kullback-Leibler distance} between two probability mass functions $p(x)$ and $q(x)$ is defined as

\begin{equation}
D\left(p \;\middle\vert\vert\; q\right) = \sum\limits_{x \in \mathcal{X}} p(x)~\log_2~\frac{p(x)}{q(x)}
\end{equation}

which is always non-negative and is zero if and only if the probabilities $p$ and $q$ are equal.

\section{Efficient representation of the states of a population}
\label{sec:mettyp}

In a population of $n$ individuals, where each of them can take $2$ states, the number of possible states that the population can take is
$2^n$. Let $s_1, \dots, s_n$ be the states of individuals $1, \dots, n$, where $s_i \in \mathcal{S}$ is the state of individual $i$, with
$1 \leq i \leq n$. The probability of the population to be in state $s_1,\dots,s_n$ given some environmental conditions $e$ is given
by $p(s_1, \dots, s_n|e) = p(s_1|e)\dots p(s_n|e)$. Considering that conditional probabilities among individuals of the same species
are equal (see Eq. \ref{eq:condprobx} and \ref{eq:condproby}), then $p(s_1, \dots, s_n|e)$ depend solely on the number of occurrences of
each state in $\mathcal{S}$. For instance, if $\mathcal{S} = \{0,1\}$, then $p(0,0,1|e) = p(0,1,0|e) = p(1,0,0|e) = p(0|e)^2 p(1|e)^1$.
In this way, the number of states of the population grows linearly with population size (see \cite{Cover2002} for a proof). Below, we show
how to compute $\Pr(S|E)$ (where $S$ represents the state of the population as a unit) by using the fact that individuals are indistinguishable.

Let $s_1, \ldots, s_n$ (or alternatively $\textbf{s}$) be a sequence of $n$ states, where $s_i \in \mathcal{S}$ is the
state of individual $i$ of species $X$, with $1 \leq i \leq n$. The \emph{type} $P_\textbf{s}$ of a sequence $\textbf{s}$ is
a probability distribution given by $P_{\textbf{s}}(a) = N(a|\textbf{s})/n$ for all $a \in \mathcal{S}$, where $N(a|\textbf{s})$
is the number of times state $a$ occurs in the sequence $\textbf{s}$. The \emph{type class} of a type $P$ is defined as the set

\[ T(P) = \{\textbf{s} \in \mathcal{S}^n : P_{\textbf{s}} = P \} \]

and the size of $T(P)$ is the number of ways of arranging $N(s_1|\textbf{s}), \ldots, N(s_{|\mathcal{S}|}|\textbf{s})$ individuals in
a sequence, which is

\[ |T(P)| = \binom{n}{N(s_1|s_\textbf{x}), \ldots, N(s_{|\mathcal{S}|}|s_\textbf{x})} =
\frac{n!}{N(s_1|s_\textbf{x})! \times \ldots \times N(s_{|\mathcal{S}|}|s_\textbf{x})!} \]

Since $s_1, \ldots, s_n$ given some environmental conditions $e$ are drawn \emph{i.i.d} according to Eq. \ref{eq:condprobx} for species
$X$ of Eq. \ref{eq:condproby} for species $Y$, the probability of $\textbf{s}$ depends only on its type and is given by

\begin{equation}
p\left(\textbf{s} \;\middle\vert\; e\right) = |T(P_\textbf{s})|~
2^{-n \big( H(P_\textbf{s})+D(P_\textbf{s}||Pr(S_i|e)) \big)}
\label{eq:congcp}
\end{equation}

(see \cite{Cover2002} for a proof). The number of states of the random variable $S$ representing a population of $n$ individuals is given
by the cardinal of the set of types,

\def\multiset#1#2{\ensuremath{\left(\kern-.3em\left(\genfrac{}{}{0pt}{}{#1}{#2}\right)\kern-.3em\right)}}

\[ |\mathcal{P}_n| = \multiset{|\mathcal{S}|}{n} = \binom{n + |\mathcal{S}| - 1}{n} \]

which is the number of $n$-multisubsets of the set $\mathcal{S}$, \emph{i.e.} the total number of combinations of the states of
$n$ individual random variables, where each one can take any state of $\mathcal{S}$, counting permutations only once. In our case,
where $\mathcal{S} = \{ 0, 1 \}$, the number of states of the random variable $S$ is $n + 1$.

%

\section{Interpolation of conditional probabilities}
\label{sec:interpolation}

Population densities in our model are represented by a value in the range $[0,1]$, and this value is mapped to the actual number of
individuals in the population, for instance for species $X$ this value is $n = p_t \times X_t \times 15$. If $n$ is an integer, then we
proceed as explained in appendix \ref{sec:mettyp}. For other cases, let us assume the states of $\floor{n}$ individuals are represented
in a sequence $\textbf{s}^{\floor{n}}$, where each state is in $\mathcal{S}$. As our model requires conditional probabilities for
continuous sequences, we define a surrogate sequence which adds a proportion $\lambda$ of state $b \in \mathcal{S}$ to sequence
$\textbf{s}^{\floor{n}}$ as $\textbf{s}^\prime(\textbf{s}^{\floor{n}}, b, \lambda)$, which we denote $\textbf{s}^\prime$ for shortness
when the arguments can be deduced from context. Thus, we have $|\textbf{s}^{\floor{n}}| \leq |\textbf{s}^\prime| \leq |\textbf{s}^
{\floor{{n + 1}}}|$, or equivalently $\floor{n} \leq |\textbf{s}^\prime| \leq \floor{n + 1}$. 
We define the type of a sequence when adding a proportion $\lambda$ of state $b$ to the sequence $\textbf{s}^{\floor{n}}$ as

\begin{equation}
P_{\textbf{s}^\prime(\textbf{s}^{\floor{n}}, b, \lambda)}(a) = \left\{
	\begin{array}{rl}
		\frac{N(a|\textbf{s}^{\floor{n}}) + \lambda}{\floor{n} + \lambda} &\mbox{$a \in \mathcal{S}, a = b$} \\ \\
		\frac{N(a|\textbf{s}^{\floor{n}})}{\floor{n} + \lambda} &\mbox{$a \in \mathcal{S}, a \neq b$} 
	\end{array}
	\right.
	\label{eq:typeinterpolation}
\end{equation}

For $\lambda = 0$ one has $P_{\textbf{s}^\prime} = P_{\textbf{s}^{\floor{n}}}$, and for $\lambda = 1$ one has $P_{\textbf{s}^\prime} =
P_{\textbf{s}^{\floor{n + 1}}}$. 

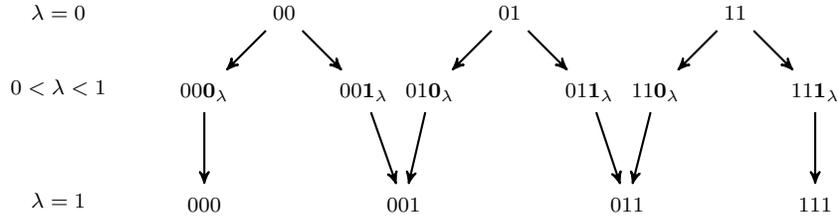
\begin{figure}[ht]
\centering
  \begin{tikzpicture}
    [->,>=stealth',shorten >=1pt,auto,node distance=1.5cm,
    thick,main node/.style={font=\sffamily\scriptsize\bfseries}]

    \node[main node] (1) [] {$01$};
    \node[main node] (2) [left of=1,node distance=3cm] {$00$};
    \node[main node] (3) [right of=1,node distance=3cm] {$11$};
    \node[main node] (4) [below left of=2] {$00\mathbf{0_\lambda}$};
    \node[main node] (5) [below right of=2] {$00\mathbf{1_\lambda}$};
    \node[main node] (6) [below left of=1] {$01\mathbf{0_\lambda}$};
    \node[main node] (7) [below right of=1] {$01\mathbf{1_\lambda}$};
    \node[main node] (8) [below left of=3] {$11\mathbf{0_\lambda}$};
    \node[main node] (9) [below right of=3] {$11\mathbf{1_\lambda}$};
    \node[main node] (10) [below of=4] {$000$};
    \node[main node] (11) [below of=6] {};
    \node[main node] (12) [below of=7] {};
    \node[main node] (13) [below of=9] {$111$};
    \node[main node] (14) [left of=2,node distance=3cm] {$\lambda = 0$};
    \node[main node] (15) [below of=14,node distance=1cm] {$0 < \lambda < 1$};
    \node[main node] (16) [below of=15] {$\lambda = 1$};
    \node[main node] (110) [left of=11,node distance=0.35cm] {$001$};
    \node[main node] (120) [right of=12,node distance=0.5cm] {$011$};


    \path[every node/.style={font=\sffamily\small}]
      (2) edge node {} (4)
      (2) edge node {} (5)
      (1) edge node {} (6)
      (1) edge node {} (7)
      (3) edge node {} (8)
      (3) edge node {} (9)
      (4) edge node {} (10)
      (5) edge node {} (110)
      (6) edge node {} (110)
      (7) edge node {} (120)
      (8) edge node {} (120)
      (9) edge node {} (13)
      ;
  \end{tikzpicture}
  \caption{States of a population in the transition from two to three individuals. We denote a sequence $\mathbf{s^\prime}(\mathbf{s}, b, \lambda)$ as
			$\mathbf{s}\mathbf{b}_\lambda$ for shortness. For instance, ($00$,$1$,$\lambda$) is denoted $00\mathbf{1}_\lambda$.}
  \label{fig:interpolation}
\end{figure}

We should note that when $0 < \lambda < 1$, the total number of states of the population is $|\mathcal{S}|\times|\mathcal{P}_n|$.
Let us illustrate the states of a population when $0 < \lambda < 1$ for a population that consists of two individuals, where each
individual can be in a state $0$ or $1$. The possible states of this population are $3$: $00$, $01$, and $11$ (with type size $1$,
$2$, $1$, respectively). Now, if we add a proportion $0 < \lambda < 1$ of a state to each possible state of the population, then we
multiply by $|\mathcal{S}|$ the number of states (before shrinking when $\lambda = 1$, see Fig. \ref{fig:interpolation}). 

We could consider the type size of each sequence by 

\[ |T(P_{\textbf{s}^\prime(\textbf{s}^{\floor{n}}, b, \lambda)})| \coloneqq
\binom{\floor{n} + \lambda}{N(s_1|s_\textbf{x}), \ldots, N(s_b|s_\textbf{x}) + \lambda, \ldots, N(s_{|\mathcal{S}|}|s_\textbf{x})} \]

where the factorial is approximated by using the gamma function $\Gamma \left(x\right) = \int\limits_0^\infty {t^{x - 1} e^{ -t} dt}$,
with $(\floor{n} + \lambda)! = \Gamma(\floor{n} + \lambda + 1)$. However, when counting the unique ways of arranging the states,
for instance, $001_\lambda$ and $010_\lambda$, we would not be considering the overlap between these two states. In other words,
we would be counting more than once some sequences. This depends on the sequences and the value of $\lambda$, for example $000_\lambda$
and $001_\lambda$ fully overlap when $\lambda = 0$, but there is no overlap when $\lambda = 1$. On the other hand, $001_\lambda$ and
$010_\lambda$ do not overlap when $\lambda = 0$, but fully overlap when $\lambda = 1$.

%

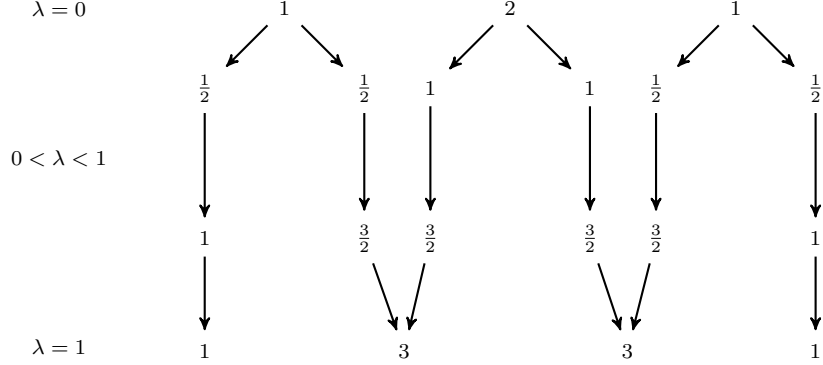
\begin{figure}[ht]
\centering
  \begin{tikzpicture}
    [->,>=stealth',shorten >=1pt,auto,node distance=1.5cm,
    thick,main node/.style={font=\sffamily\scriptsize\bfseries}]

    \node[main node] (1) [] {$2$};
    \node[main node] (2) [left of=1,node distance=3cm] {$1$};
    \node[main node] (3) [right of=1,node distance=3cm] {$1$};
    \node[main node] (4) [below left of=2] {$\frac{1}{2}$};
    \node[main node] (5) [below right of=2] {$\frac{1}{2}$};
    \node[main node] (6) [below left of=1] {$1$};
    \node[main node] (7) [below right of=1] {$1$};
    \node[main node] (8) [below left of=3] {$\frac{1}{2}$};
    \node[main node] (9) [below right of=3] {$\frac{1}{2}$};
    \node[main node] (50) [below of=4,node distance=2cm] {$1$};
    \node[main node] (51) [below of=5,node distance=2cm] {$\frac{3}{2}$};
    \node[main node] (52) [below of=6,node distance=2cm] {$\frac{3}{2}$};
    \node[main node] (53) [below of=7,node distance=2cm] {$\frac{3}{2}$};
    \node[main node] (54) [below of=8,node distance=2cm] {$\frac{3}{2}$};
    \node[main node] (55) [below of=9,node distance=2cm] {$1$};
    \node[main node] (10) [below of=50] {$1$};
    \node[main node] (11) [below of=52] {};
    \node[main node] (12) [below of=53] {};
    \node[main node] (13) [below of=55] {$1$};
    \node[main node] (14) [left of=2,node distance=3cm] {$\lambda = 0$};
    \node[main node] (15) [below of=14,node distance=2cm] {$0 < \lambda < 1$};
    \node[main node] (16) [below of=15,node distance=2.5cm] {$\lambda = 1$};
    \node[main node] (80) [left of=11,node distance=0.35cm] {$3$};
    \node[main node] (81) [right of=12,node distance=0.5cm] {$3$};


    \path[every node/.style={font=\sffamily\small}]
      (2) edge node {} (4)
      (2) edge node {} (5)
      (1) edge node {} (6)
      (1) edge node {} (7)
      (3) edge node {} (8)
      (3) edge node {} (9)
      (4) edge node {} (50)
      (5) edge node {} (51)
      (6) edge node {} (52)
      (7) edge node {} (53)
      (8) edge node {} (54)
      (9) edge node {} (55)
      (50) edge node {} (10)
      (51) edge node {} (80)
      (52) edge node {} (80)
      (53) edge node {} (81)
      (54) edge node {} (81)
      (55) edge node {} (13)
      ;
  \end{tikzpicture}
  \caption{Sizes of types for sequences in the transition from two to three individuals in a population. The sizes are corresponded
			with the sequences shown in Fig. \ref{fig:interpolation}, and the values when $0 < \lambda < 1$ show how they change from
			$\lambda = 0$ to $\lambda = 1$.}
  \label{fig:pascal}
\end{figure}

Then, some sequences (such as $001_\lambda$ and $010_\lambda$) are always counted twice, independently of the value of $\lambda$:
they either overlap with one sequence or the other (this is because we are considering $|\mathcal{S}| = 2$). However, the sequences
$000_\lambda$ and $111_\lambda$ (those such that $|T(P_{000_\lambda})| = T(P_{111_\lambda})| = 1$) are counted twice when $\lambda = 0$,
but only once when $\lambda = 1$. In Fig. \ref{fig:pascal} we show some values of the sizes of types as $\lambda$ increases.

Taking this into account, we approximate the conditional probability by


\begin{equation}
p(\textbf{s}^\prime(\textbf{s}^{\floor{n}}, b, \lambda) | e) = \left\{
	\begin{array}{rl}
		\frac{1 + \lambda}{|\mathcal{S}|}~
		2^{-(\floor{n} + \lambda) \big( H(P_{\textbf{s}^\prime})+D(P_{\textbf{s}^\prime}||Pr(S^\prime|E)) \big)}
		&\mbox{if $|T(P_{\textbf{s}^\prime})| = 1$} \\ \\
		\frac{|T(P_{\textbf{s}^\prime})|}{|\mathcal{S}|}~
		2^{-(\floor{n} + \lambda) \big( H(P_{\textbf{s}^\prime})+D(P_{\textbf{s}^\prime}||p(S^\prime|E)) \big)}
		&\mbox{otherwise}
	\end{array}
	\right.
	\label{eq:cpinterpolation}
\end{equation}

\section{Parameters sensitivity and results generality}
\label{sec:params}

The parameter settings used in Sec. \ref{sec:results} were specially chosen to show the transition from cooperation to antagonism in
species sharing environmental information. These were $\alpha = 1.05, N = M = 15$, together with the conditional probabilities shown
in Eq. \ref{eq:condprobx} and Eq. \ref{eq:condproby}. Here, we analyse the sensitivity of the parameters by introducing changes in each
one of them and showing how this affects the results. Instead of computing the dominant strategies for the same subset of
$\mathcal{X} \times \mathcal{Y} \times \mathcal{R}$ (as we have done in Sec. \ref{sec:results}), we show results for a fixed value of
resources, $R = 1.8$, which clearly shows all the possible volumes (see Fig. \ref{fig:r1.8}).

\begin{table}[ht]
	\centering
	\renewcommand\arraystretch{1.2}
	\begin{tabular}{| c | l |}
		\cline{1-2}
		\textbf{Parameter} & \textbf{Description} \\
		\cline{1-2}
		\multirow{2}{*}{$\Pr\left(S_{x_i}\;\middle\vert\;E\right)$} & \small{This conditional probability defines the amount of information that} \\ 
		& \small{an individual cell $i$ of species $X$ captures from sensing the environment.} \\ 
		\cline{1-2}
		\multirow{2}{*}{$\Pr\left(S_{y_j}\;\middle\vert\;E\right)$} & \small{This conditional probability defines the amount of information that} \\ 
		& \small{an individual cell $j$ of species $Y$ captures from sensing the environment.} \\ 
		\cline{1-2}
		\multirow{2}{*}{$N$} & \multirow{2}{*}{\small{Carrying capacity of the population of species $X$.}} \\ & \\
		\cline{1-2}
		\multirow{2}{*}{$M$} & \multirow{2}{*}{\small{Carrying capacity of the population of species $Y$.}} \\ & \\
		\cline{1-2}
		\multirow{2}{*}{$\alpha$} & \multirow{2}{*}{\small{Growth rate of resources.}} \\ & \\
		\cline{1-2}
		\end{tabular}
	\caption{List of parameters used by the model with their description.}
	\label{tab:params}
\end{table}

\begin{center}
\begin{figure}[ht]
	\centering
	\noindent\begin{minipage}[b]{.08\textwidth}
		\rotatebox{90}{\tiny{~~~~~~~~~~$X$ population density}}
	\end{minipage}%
	\begin{minipage}[b]{.23\textwidth}
		\includegraphics[scale=0.6]{a_slices_180.pdf}\\
		\vspace*{0.1cm}\hspace*{0.3cm}{\tiny{$Y$ population density}}
		\subcaption{$\alpha = 1.05$}
		\label{fig:alpha_def_r1.8}
	\end{minipage}%
	\begin{minipage}[b]{.23\textwidth}
		\includegraphics[scale=0.6]{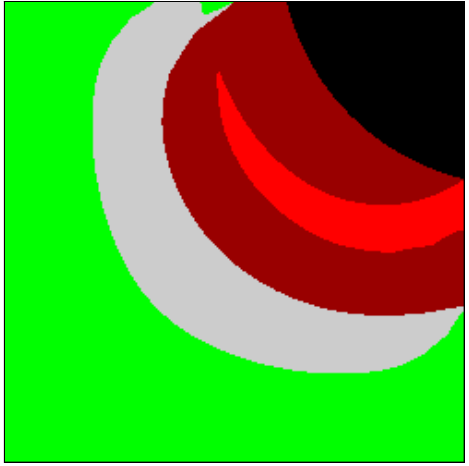}\\
		\vspace*{0.1cm}\hspace*{0.3cm}{\tiny{$Y$ population density}}
		\subcaption{$\alpha = 1.25$}
		\label{fig:alpha_hi_r1.8}
	\end{minipage}%
	\begin{minipage}[b]{.23\textwidth}
		\includegraphics[scale=0.6]{a_slices_180.pdf}\\
		\vspace*{0.1cm}\hspace*{0.3cm}{\tiny{$Y$ population density}}
		\fontsize{1pt}{3pt}
		\subcaption{$\Pr(S_{X_i}|E)$ (\ref{eq:condprobx}), $\Pr(S_{Y_j}|E)$ (\ref{eq:condproby})}
		\label{fig:diff_def_r1.8}
	\end{minipage}%
	\begin{minipage}[b]{.23\textwidth}
		\includegraphics[scale=0.6]{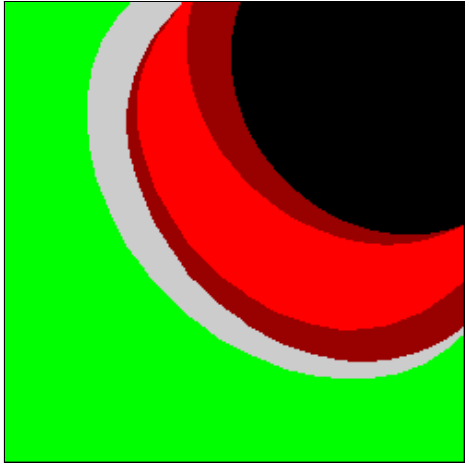}\\
		\vspace*{0.1cm}\hspace*{0.3cm}{\tiny{$Y$ population density}}
		\subcaption{$\Pr(S_{X_i}|E)$ (\ref{eq:condprobxmod}), $\Pr(S_{Y_j}|E)$ (\ref{eq:condprobymod})}
		\label{fig:diff_hi_r1.8}
	\end{minipage}%
	\caption{(see in colour) We analyse the payoff matrix from $250^2$ initial values of population densities of species $X$ and $Y$
			uniformly distributed in $\mathcal{X}\times\mathcal{Y}$, with the amount of resources $R = 1.8$ fixed. See Fig.
			\ref{fig:gameanalysis} for an explanation of what colors represent. Each subcaption shows the parameter whose effect is being
			exemplified.
			}
	\label{fig:sensitivityanalysis}
\end{figure}
\end{center}

In Table \ref{tab:params}, we show the used parameters by the model with their descriptions. First, let us consider parameter $\alpha$,
the growth rate of resources. In Fig. \ref{fig:alpha_hi_r1.8}, we show the results we obtain when we change to $\alpha = 1.25$, instead
of its original value, $\alpha = 1.05$, whose results are shown in Fig. \ref{fig:alpha_def_r1.8}. This change extends the volume where
sharing information is weakly dominant such that it includes initial values with relatively lower resources, while the other volumes will be shifted in
such a way that they are composed of initial values with lower values for resources. Additionally, the amount of initial values composing
the other volumes is smaller. The reason for this is that, since resources grow at a higher rate, higher populations can be supported,
and thus the regime in which antagonism is dominant is reduced.


\begin{figure}[H]
	\begin{center}
	\begin{tikzpicture}[gnuplot]
	\gpcolor{gp lt color border}
	\gpsetlinetype{gp lt border}
	\gpsetlinewidth{1.00}
	\draw[gp path] (1.504,0.985)--(1.684,0.985);
	\draw[gp path] (11.947,0.985)--(11.767,0.985);
	\node[gp node right] at (1.320,0.985) { 0};
	\draw[gp path] (1.504,2.746)--(1.684,2.746);
	\draw[gp path] (11.947,2.746)--(11.767,2.746);
	\node[gp node right] at (1.320,2.746) { 0.5};
	\draw[gp path] (1.504,4.507)--(1.684,4.507);
	\draw[gp path] (11.947,4.507)--(11.767,4.507);
	\node[gp node right] at (1.320,4.507) { 1};
	\draw[gp path] (1.504,6.268)--(1.684,6.268);
	\draw[gp path] (11.947,6.268)--(11.767,6.268);
	\node[gp node right] at (1.320,6.268) { 1.5};
	\draw[gp path] (1.504,8.029)--(1.684,8.029);
	\draw[gp path] (11.947,8.029)--(11.767,8.029);
	\node[gp node right] at (1.320,8.029) { 2};
	\draw[gp path] (1.504,0.985)--(1.504,1.165);
	\draw[gp path] (1.504,8.381)--(1.504,8.201);
	\node[gp node center] at (1.504,0.677) { 0};
	\draw[gp path] (2.896,0.985)--(2.896,1.165);
	\draw[gp path] (2.896,8.381)--(2.896,8.201);
	\node[gp node center] at (2.896,0.677) { 2};
	\draw[gp path] (4.289,0.985)--(4.289,1.165);
	\draw[gp path] (4.289,8.381)--(4.289,8.201);
	\node[gp node center] at (4.289,0.677) { 4};
	\draw[gp path] (5.681,0.985)--(5.681,1.165);
	\draw[gp path] (5.681,8.381)--(5.681,8.201);
	\node[gp node center] at (5.681,0.677) { 6};
	\draw[gp path] (7.074,0.985)--(7.074,1.165);
	\draw[gp path] (7.074,8.381)--(7.074,8.201);
	\node[gp node center] at (7.074,0.677) { 8};
	\draw[gp path] (8.466,0.985)--(8.466,1.165);
	\draw[gp path] (8.466,8.381)--(8.466,8.201);
	\node[gp node center] at (8.466,0.677) { 10};
	\draw[gp path] (9.858,0.985)--(9.858,1.165);
	\draw[gp path] (9.858,8.381)--(9.858,8.201);
	\node[gp node center] at (9.858,0.677) { 12};
	\draw[gp path] (11.251,0.985)--(11.251,1.165);
	\draw[gp path] (11.251,8.381)--(11.251,8.201);
	\node[gp node center] at (11.251,0.677) { 14};
	\draw[gp path] (1.504,8.381)--(1.504,0.985)--(11.947,0.985)--(11.947,8.381)--cycle;
	\node[gp node center,rotate=-270] at (0.246,4.683) {environmental information (in bits)};
	\node[gp node center] at (6.725,0.215) {population size of species X (n) and Y (m)};
	\node[gp node right] at (11.2,5) {\small{$H(E)$}};
	\gpcolor{gp lt color 0}
	\gpsetlinetype{gp lt plot 0}
	\draw[gp path] (11.25,5)--(11.65,5);
	\draw[gp path] (1.504,8.029)--(2.200,8.029)--(2.896,8.029)--(3.593,8.029)--(4.289,8.029)%
	  --(4.985,8.029)--(5.681,8.029)--(6.377,8.029)--(7.074,8.029)--(7.770,8.029)--(8.466,8.029)%
	  --(9.162,8.029)--(9.858,8.029)--(10.555,8.029)--(11.251,8.029)--(11.947,8.029);
	\gpcolor{gp lt color border}
	\node[gp node right] at (11.2,4.5) {\small{$I(E;S_{X_i}), I(E;S_{Y_j})$}};
	\gpcolor{gp lt color 1}
	\gpsetlinetype{gp lt plot 1}
	\draw[gp path] (11.25,4.5)--(11.65,4.5);
	\draw[gp path] (1.504,0.985)--(2.200,2.358)--(2.896,2.358)--(3.593,2.358)--(4.289,2.358)%
	  --(4.985,2.358)--(5.681,2.358)--(6.377,2.358)--(7.074,2.358)--(7.770,2.358)--(8.466,2.358)%
	  --(9.162,2.358)--(9.858,2.358)--(10.555,2.358)--(11.251,2.358)--(11.947,2.358);
	\gpcolor{gp lt color border}
	\node[gp node right] at (11.2,4) {\small{$I(E;S_{X_1},\dots,S_{X_n}),I(E;S_{Y_1},\dots,S_{Y_m})$}};
	\gpcolor{gp lt color 2}
	\gpsetlinetype{gp lt plot 2}
	\draw[gp path] (11.25,4)--(11.65,4);
	\draw[gp path] (1.504,0.985)--(2.200,2.358)--(2.896,3.197)--(3.593,3.806)--(4.289,4.281)%
	  --(4.985,4.669)--(5.681,4.995)--(6.377,5.276)--(7.074,5.521)--(7.770,5.737)--(8.466,5.930)%
	  --(9.162,6.102)--(9.858,6.258)--(10.555,6.398)--(11.251,6.526)--(11.947,6.643);
	\gpcolor{gp lt color border}
	\node[gp node right] at (11.2,3.5) {\small{$I(E;S_{X_1},\dots,S_{X_n},S_{Y_1},\dots,S_{Y_m})$}};
	\gpcolor{gp lt color 3}
	\gpsetlinetype{gp lt plot 3}
	\draw[gp path] (11.25,3.5)--(11.65,3.5);
	\draw[gp path] (1.504,0.985)--(2.200,3.197)--(2.896,4.281)--(3.593,4.995)--(4.289,5.521)%
	  --(4.985,5.930)--(5.681,6.258)--(6.377,6.526)--(7.074,6.749)--(7.770,6.934)--(8.466,7.090)%
	  --(9.162,7.221)--(9.858,7.331)--(10.555,7.425)--(11.251,7.505)--(11.947,7.574);
	\gpcolor{gp lt color border}
	\gpsetlinetype{gp lt border}
	\draw[gp path] (1.504,8.381)--(1.504,0.985)--(11.947,0.985)--(11.947,8.381)--cycle;
	\gpdefrectangularnode{gp plot 1}{\pgfpoint{1.504cm}{0.985cm}}{\pgfpoint{11.947cm}{8.381cm}}
	\end{tikzpicture}
	\caption{Total amount of environmental information for different scenarios (using conditional probabilities \ref{eq:condprobxmod} and
			\ref{eq:condprobymod}): $I(E;S_{X_i})$ and $I(E;S_{Y_j})$ correspond to the case where an individual cell $i$ of species $X$
			and an individual cell $j$ of species $Y$ acquire information from their sensors only, respectively. $I(E;S_{X_1},\dots,S_{X_n})$
			is the total amount of information of each cell of species $X$ when $n$ cells communicate; in the same way
			$I(E;S_{Y_1},\dots,S_{Y_m})$ is the total amount of information of each cell of species $Y$ when $m$ cells communicate.
			$I(E;S_{X_1},\dots,S_{X_n},S_{Y_1},\dots,S_{Y_m})$ is the total amount of environmental information each cell of both population
			have when $n$ cells of species $X$ and $m$ cells of species $Y$ communicate. Finally, $H(E)$ is the uncertainty of the environment.
			}
	\label{fig:envinfoch}
	\end{center}
\end{figure}
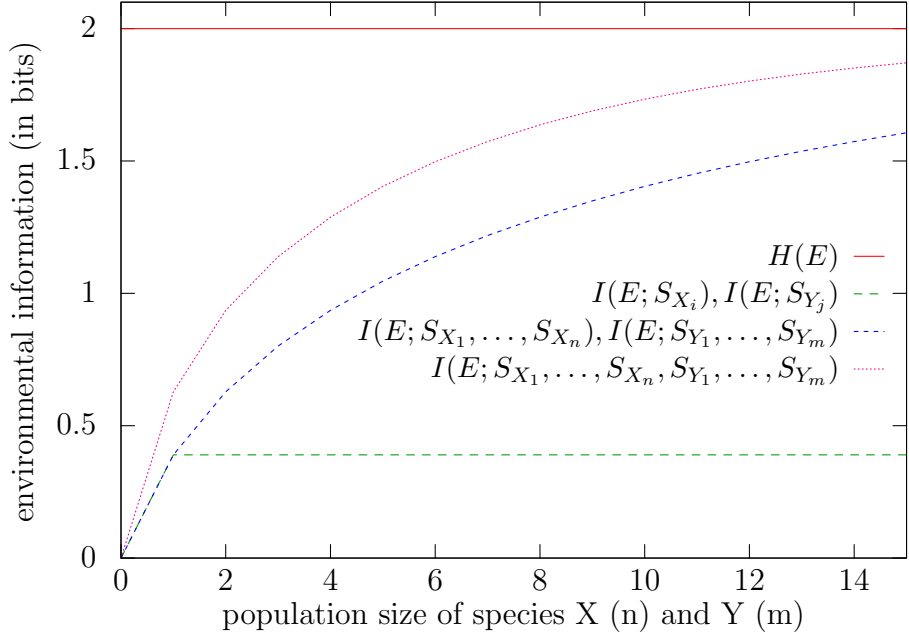

Let us consider now the conditional probabilities $\Pr(S_{X_i}|E)$ and $\Pr(S_{Y_j}|E)$. They determine how much environmental information
each individual captures, $I(E;S_{X_i}) = I(E;S_{Y_j}) = 0.39016$ bits, which is a low amount of the total environmental uncertainty,
$H(E) = 2$ bits. Together with $N$ and $M$, the conditional probabilities determine the different curves shown in Fig. \ref{fig:envinfo}.
A property of the defined conditional probabilities is that populations achieve a fast increase in environmental information when
population densities are low, with small increases for high population densities. Lower values of $I(E;S_{X_i})$ and $I(E;S_{Y_j})$ would
have a slower growth for low population densities, and species will not acquire high amounts of environmental information for high
population densities (see Fig. \ref{fig:envinfoch} for an example).

Therefore, the two conditional probabilities together with $N$ and $M$ will define the total environmental information when species
communicate. This amount is \\ $I(E;S_{X_1},\dots,S_{X_n},S_{Y_1},\dots,S_{Y_m})$ (see the curve in Fig. \ref{fig:envinfo} and in Fig.
\ref{fig:envinfoch}), and the gain in environmental information for a species receiving the shared information from the other species
(assuming equal population densities) is the difference between the two mentioned curves. This difference is high for our chosen
conditional probabilities since the species were meant to capture different aspects of the environment (see Fig. \ref{fig:envinfodiag} and
\ref{fig:envinfo}), but if the aspects of the environment that species capture intersect, then this difference decreases (see, for
instance, Fig. \ref{fig:envinfoch}).

In Fig. \ref{fig:diff_def_r1.8} and Fig. \ref{fig:diff_hi_r1.8}, we show how changing the conditional probabilities $\Pr(S_{x_i}|E)$
and $\Pr(S_{Y_j}|E)$ affect the results. Instead of the probabilities defined in Eq. \ref{eq:condprobx} and in Eq. \ref{eq:condproby},
we used the ones defined in Eq. \ref{eq:condprobxmod} and in Eq. \ref{eq:condprobymod}, where the intersection of the acquired environmental
information from sensors between two individuals of different species is $I(S_{X_i};S_{Y_j}) = 0.151452$ bits, while the amount of
environmental information each individual captures is roughly the same as before, $I(E;S_{X_i}) = I(E;S_{Y_j}) = 0.389767$ bits. However,
how the amount of environmental information changes in relation to the population densities is different from the original definitions,
as shown in Fig. \ref{fig:envinfoch}.

In this case, a species by itself is able to capture more information about the environment (roughly $1.5$ bits while before it was close
to $1$ bit). Therefore, higher populations will consume more resources than before even when they do not exchange information, and, as a
consequence, the area where species always get extinct (the black area) now includes initial conditions where before they could survive
by not sharing information (see Fig. \ref{fig:diff_def_r1.8} and \ref{fig:diff_hi_r1.8}, initial conditions that were dark red are
now black). Similarly, the area where not sharing information is strictly dominant now also contains initial values that originally were
classified as not sharing weakly dominant. The reason for this is the same as mentioned: species by themselves capture more information
than before, and therefore the consumption of resources is higher even when species share information only in the first time-step (such
situations are exemplified in Table \ref{tab:nwd_1}). Another consequence of the defined conditional probabilities is that the area where
species cooperate was slightly extended. This happens because the amounts of consumed resources for high population densities decreased
in comparison to our original parameters, since high population densities now possess less environmental information. 

\begin{figure*}[ht]
	\centering
	\begin{minipage}[b]{.49\textwidth}
		\begin{equation}
		\Pr\left(S_{X_i}\;\middle\vert\;E\right) \coloneqq 
		\bordermatrix{~ & s_1 & s_2 \cr
		              e_1 & 0.95 & 0.05 \cr
		              e_2 & 0.65 & 0.35 \cr
		              e_3 & 0.35 & 0.65 \cr
		              e_4 & 0.05 & 0.95 \cr}
		\label{eq:condprobxmod}
		\end{equation}
	\end{minipage}\hfill
	\begin{minipage}[b]{.49\textwidth}
		\begin{equation}
		\Pr\left(S_{Y_i}\;\middle\vert\;E\right) \coloneqq 
		\bordermatrix{~ & s_1 & s_2 \cr
		              e_1 & 0.05 & 0.95 \cr
		              e_2 & 0.35 & 0.65 \cr
		              e_3 & 0.65 & 0.35 \cr
		              e_4 & 0.95 & 0.05 \cr}
		\label{eq:condprobymod}
		\end{equation}
	\end{minipage}
\end{figure*}

To summarise, changes in the parameters can extend or reduce the initial conditions where species cooperate, and they also can extend or
reduce the initial conditions where species get extinct independently of their strategies. These changes also affect the initial
conditions where there is no dominant strategy, where not sharing is strictly dominant, and where not sharing is weakly dominant.
However, in the results there is always a transition from cooperative strategies to antagonistic strategies, and the parameters we
chose are ones that clearly show it.

\section{Different dynamics for resources}
\label{sec:other_resource_dynamics}

In the model, we defined a dynamics for resources such that they grow at a rate of $\alpha$ unless they are exhausted, in which case
they remain in that state. Here, we consider a periodic replenishment of resources which is independent of the current amount available
for bacteria. The dynamics are shown in Eq. \ref{eq:resources_replenishment}.

\begin{equation}
	R_{t + 1} \coloneqq R_t - (X_t + Y_t) + \beta
	\label{eq:resources_replenishment}
\end{equation}

In this equation, resources are consumed proportionally to the sum of the population densities, and they are replenished by an amount
$\beta$. We would like to test whether this change affects the observed transition from cooperation to antagonism in information
exchange. For this purpose, we re-run the experiments of Sec. \ref{sec:results}, but instead of using Eq. \ref{eq:resources} in the
model, we use the one defined in this section, Eq. \ref{eq:resources_replenishment}, with a value $\beta = 0.05$ (this value has
been chosen particularly to show the transition of regimes from cooperation to antagonism).

\begin{center}
\begin{figure}[ht]
	\centering
	\noindent\begin{minipage}[b]{.05\textwidth}
		\rotatebox{90}{\tiny{~~~~~~~~~~$X$ population density}}
	\end{minipage}%
	\noindent\begin{minipage}[b]{.19\textwidth}
		\includegraphics[scale=0.6]{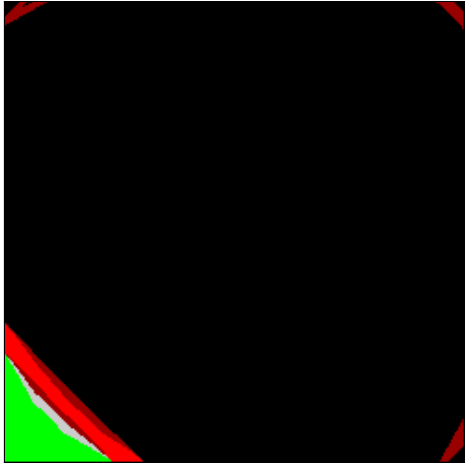}
		\vspace*{0.1cm}\hspace*{0.3cm}{\tiny{$Y$ population density}}
		\subcaption{$R = 0.4$}
		\label{fig:b_r0.4}
	\end{minipage}%
	\noindent\begin{minipage}[b]{.19\textwidth}
		\includegraphics[scale=0.6]{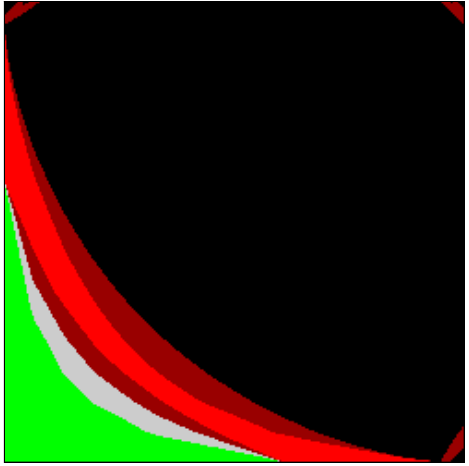}
		\vspace*{0.1cm}\hspace*{0.3cm}{\tiny{$Y$ population density}}
		\subcaption{$R = 0.9$}
		\label{fig:b_r0.9}
	\end{minipage}%
	\noindent\begin{minipage}[b]{.19\textwidth}
		\includegraphics[scale=0.6]{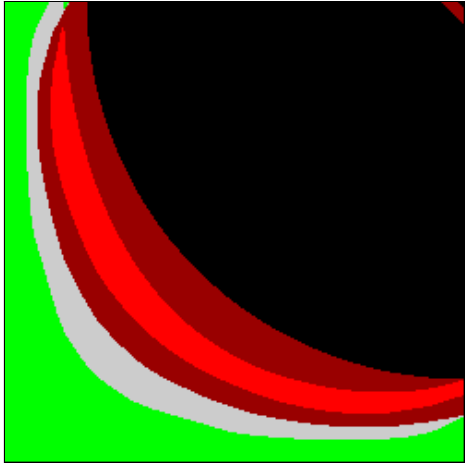}
		\vspace*{0.1cm}\hspace*{0.3cm}{\tiny{$Y$ population density}}
		\subcaption{$R = 1.2$}
		\label{fig:b_r1.2}
	\end{minipage}%
	\noindent\begin{minipage}[b]{.19\textwidth}
		\includegraphics[scale=0.6]{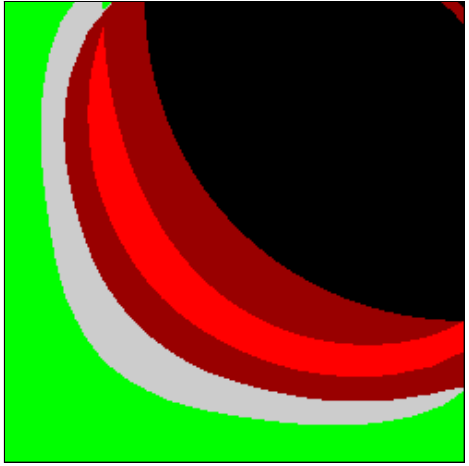}
		\vspace*{0.1cm}\hspace*{0.3cm}{\tiny{$Y$ population density}}
		\subcaption{$R = 1.4$}
		\label{fig:b_r1.4}
	\end{minipage}%
	\noindent\begin{minipage}[b]{.19\textwidth}
		\includegraphics[scale=0.6]{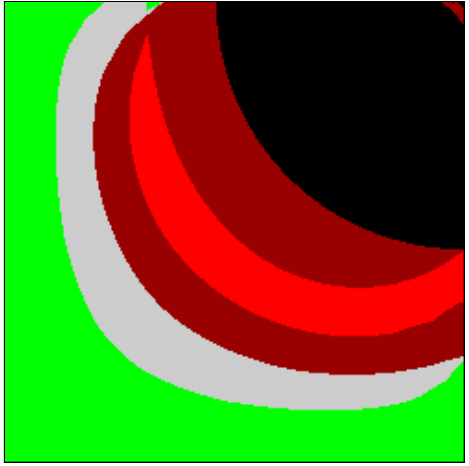}
		\vspace*{0.1cm}\hspace*{0.3cm}{\tiny{$Y$ population density}}
		\subcaption{$R = 1.6$}
		\label{fig:b_r1.6}
	\end{minipage}%
	\vspace{0.35cm}
	\centering
	\noindent\begin{minipage}[b]{.05\textwidth}
		\rotatebox{90}{\tiny{~~~~~~~~~~$X$ population density}}
	\end{minipage}%
	\begin{minipage}[b]{.19\textwidth}
		\includegraphics[scale=0.6]{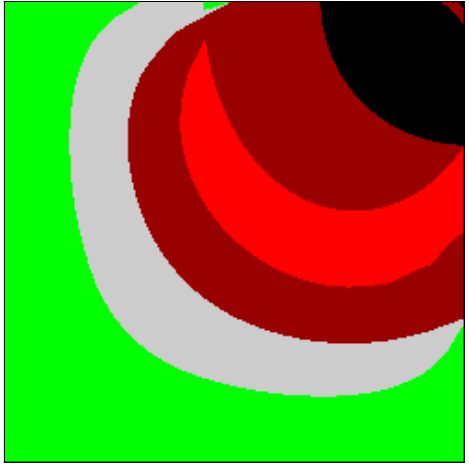}
		\vspace*{0.1cm}\hspace*{0.3cm}{\tiny{$Y$ population density}}
		\subcaption{$R = 1.8$}
		\label{fig:b_r1.8}
	\end{minipage}%
	\begin{minipage}[b]{.19\textwidth}
		\includegraphics[scale=0.6]{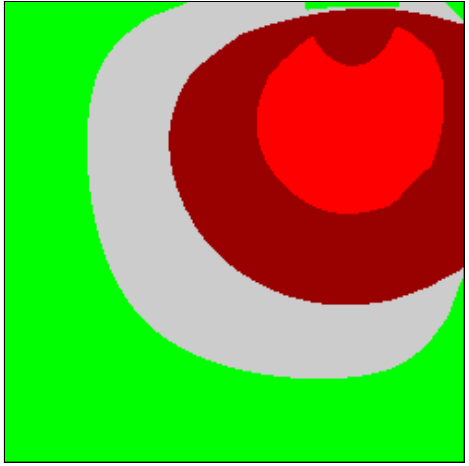}
		\vspace*{0.1cm}\hspace*{0.3cm}{\tiny{$Y$ population density}}
		\subcaption{$R = 2.0$}
		\label{fig:b_r2.0}
	\end{minipage}%
	\begin{minipage}[b]{.19\textwidth}
		\includegraphics[scale=0.6]{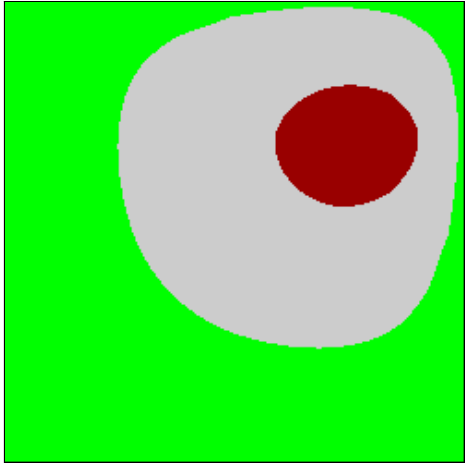}
		\vspace*{0.1cm}\hspace*{0.3cm}{\tiny{$Y$ population density}}
		\subcaption{$R = 2.3$}
		\label{fig:b_r2.3}
	\end{minipage}%
	\begin{minipage}[b]{.19\textwidth}
		\includegraphics[scale=0.6]{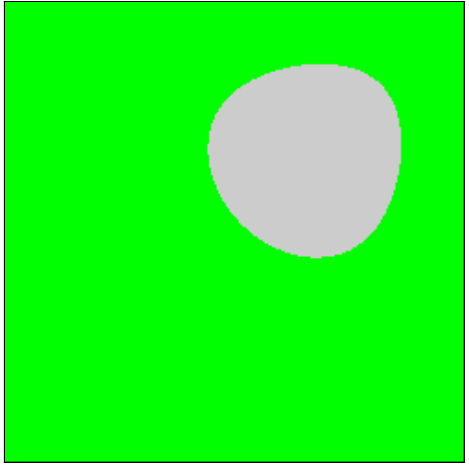}
		\vspace*{0.1cm}\hspace*{0.3cm}{\tiny{$Y$ population density}}
		\subcaption{$R = 2.8$}
		\label{fig:b_r2.8}
	\end{minipage}%
	\begin{minipage}[b]{.19\textwidth}
		\includegraphics[scale=0.6]{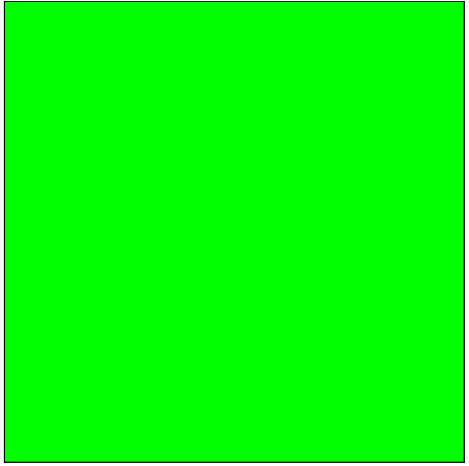}
		\vspace*{0.1cm}\hspace*{0.3cm}{\tiny{$Y$ population density}}
		\subcaption{$R = 3.0$}
		\label{fig:b_r3.0}
	\end{minipage}%
	\caption{(see in colour) We analyse the payoff matrix from $250^2$ initial values of population densities of species $X$ and
			$Y$ uniformly distributed in $\mathcal{X}\times\mathcal{Y}$, with the amount of resources $R$ fixed. Green points corresponds
			to sharing information being a weakly dominant strategy for species $X$. Grey points corresponds to initial values where there
			is no dominant strategy for species $X$. Red points represent values for which not sharing information is strictly
			dominant for species $X$. Dark-red points corresponds to values where not sharing information is weakly dominant for
			species $X$. Finally, black points represent values where species $X$ goes extinct regardless of its strategy.}
	\label{fig:b_gameanalysis}
\end{figure}
\end{center}

The plots shown in Fig. \ref{fig:b_gameanalysis} are very similar to the ones shown in Fig. \ref{fig:gameanalysis}, the latter corresponding
to the figures obtained from the original simulations.
The difference that stands out between the two experiments happens for population densities
close to their carrying capacity. In the original simulations, such population densities went extinct independently of their actions. In the
current consideration, not sharing information is weakly dominant. Before, large populations would exhaust resources, and, since they were
never replenished, both species would die out. Now, resources are may be fully consumed, but they would be replenished in any case.

\bibliographystyle{abbrv}
\bibliography{manuscript}

\end{document}